\documentclass[10pt,aps,prd,twocolumn,showpacs,superscriptaddress,eqsecnum,longbibliography,nofootinbib]{revtex4-2}
\usepackage{mathrsfs,amsmath,amsthm,latexsym,amssymb,amsfonts,epsfig,cancel,enumerate,graphicx,subcaption,txfonts,overpic}
\usepackage{multirow}
\usepackage{xcolor}
\definecolor{navy}{RGB}{0,0,150}
\usepackage[colorlinks, linkcolor=blue, citecolor=blue, urlcolor=blue, plainpages=false, pdfstartview=FitH]{hyperref}
\usepackage{appendix}
\allowdisplaybreaks
\newcommand{\YO}{Yunnan Observatories, Chinese Academy of Sciences, Kunming 650216, People's Republic of China}
\newcommand{\GZU}{College of Physics, Guizhou University, Guiyang 550025, China}
\newcommand{\HU}{College of Physics and Science and Technology, Hebei University, Baoding 071002, China}
\begin{document}
	
	\title{Periodic orbits and gravitational waveforms around a Schwarzschild black hole with a cloud of strings embedded in perfect fluid dark matter}
	
	\author{Ziqiang Cai}
	\email{gs.zqcai24@gzu.edu.cn}
	\affiliation{\GZU}
	
	\author{Zhi Li}
	\email{lizhi@ynao.ac.cn}
	\affiliation{\YO}

	\author{Zhenglong Ban}
	\email{zlban123@163.com}
	\affiliation{\HU}
	
	\author{Qi-Qi Liang}
	\email{13985671614@163.com}
	\affiliation{\GZU}
	
	\author{Zheng-Wen Long}
	\thanks{Corresponding author}
	\email{zwlong@gzu.edu.cn}
	\affiliation{\GZU}
	
	\begin{abstract}
	In this study, we explore the dynamics of particle orbits and their corresponding gravitational wave signatures in the vicinity of a Schwarzschild black hole (BH) surrounded by a cloud of strings and embedded in a perfect fluid dark matter medium. The model is characterized by two parameters: $a$, associated with the string cloud, and $\alpha$, representing the dark matter distribution. We systematically analyze how the marginally bound orbit (MBO) and the innermost stable circular orbit (ISCO) depend on these parameters. Our findings reveal that while both the orbital radius and angular momentum increase with increasing $a$, they decrease as $\alpha$ increases; notably, the energy exhibits the opposite trend, decreasing with $a$ and increasing with $\alpha$. Furthermore, we examine periodic orbits indexed by rational numbers $q$ and the gravitational waveforms they generate. The results demonstrate that an increase in the string cloud parameter $a$ induces a significant phase delay in the waveform. Specifically, waveforms with lower values of $a$  oscillate over shorter time intervals, whereas those with higher values extend to longer time scales. These distinct features, including noticeable differences in amplitude, allow the waveforms to be clearly distinguished from those in a pure Schwarzschild spacetime.
	\end{abstract}
	
	\maketitle
	\section{Introduction}
	Among the most fascinating predictions of General Relativity (GR) is the existence of BHs, celestial entities often described as the universe's most mysterious components. The landmark gravitational wave detections of merging BHs by LIGO represent a dual triumph: they serve as a robust confirmation of GR's fundamental tenets while simultaneously providing researchers with an unprecedented tool for probing BH properties \cite{LIGOScientific:2016aoc}. This observational revolution was complemented by the EHT Collaboration's 2019 imaging of the supermassive BH in M87$^{*}$ \cite{EventHorizonTelescope:2019dse,EventHorizonTelescope:2019uob,EventHorizonTelescope:2019jan,EventHorizonTelescope:2019ths,EventHorizonTelescope:2019pgp,EventHorizonTelescope:2019ggy}, followed by the imaging of Sgr A$^*$ \cite{EventHorizonTelescope:2022wkp,EventHorizonTelescope:2022apq,EventHorizonTelescope:2022wok,EventHorizonTelescope:2022exc,EventHorizonTelescope:2022urf,EventHorizonTelescope:2022xqj}. These observations stand as a testament to the accuracy of theoretical models, carrying distinct imprints of the dynamic processes that govern the material encircling these compact objects \cite{EventHorizonTelescope:2021bee}.
	
	String theory provides a unique perspective on BHs by substituting one-dimensional strings for the traditional zero-dimensional particles. Of particular interest is the proposal that a cloud of strings can serve as a material source of gravity. Building on this idea, Letelier formulated an exact analytical model for a Schwarzschild BH surrounded by a cloud of strings \cite{Letelier:1979ej}. This seminal contribution has spurred extensive investigations into how string clouds modify the properties of various BH spacetimes \cite{Toledo:2018hav,Cardenas:2021eri,Mustafa:2021hvq,Rodrigues:2022zph,Zahid:2023csk,Sudhanshu:2024wqb}. Despite accounting for roughly 27\% of the cosmic energy density, the true nature of dark matter continues to elude physicists, underscoring the critical need for its study. Such research is indispensable for explaining the genesis of galaxies and the universe's large-scale architecture, as well as for examining dark matter's interaction with dark energy, which fuels the accelerated cosmic expansion. The Perfect Fluid Dark Matter model offers a distinctive approach to exploring dark matter's role, particularly concerning BHs. In contrast to conventional particle-based models, perfect fluid dark matter treats dark matter as a continuous, non-viscous fluid governed by specific equations of state \cite{Kiselev:2003ah}. Applications of this model have revealed how dark matter clustering around BHs \cite{Qiao:2022nic} can alter their observable properties.
	
	Although BHs are under intense theoretical and observational study, they continue to captivate researchers as some of the most intriguing objects in the universe. Einstein's GR, which interprets gravity as the curvature of spacetime, underpins our modern understanding of gravity and has successfully predicted phenomena like gravitational waves (GWs) \cite{LIGOScientific:2016aoc,LIGOScientific:2016vlm}. The direct observation of GWs has opened a unique window into the strong-gravity regime. For compact objects such as BHs, GW detection and analysis are essential to deciphering their dynamics. Current detectors, such as LISA \cite{LISA:2017pwj} and Taiji \cite{Hu:2017mde}, largely focus on comparable-mass stellar binaries. However, Extreme Mass-Ratio Inspirals (EMRIs) \cite{Hughes:2000ssa,Amaro-Seoane:2012lgq,Babak:2017tow} constitute a vital alternative source. In these systems, a stellar-mass object spirals into a Supermassive BH, emitting low-frequency GWs. These waveforms yield precise information on orbital dynamics and allow for parameterized tests of deviations from Kerr spacetime \cite{Ghosh:2024arw,Ghosh:2024het}. Critically, EMRI-generated GWs provide a wealth of data on near-horizon phenomena and the properties of dark matter envelopes around SMBHs \cite{Yue:2018vtk,Duque:2023seg,Dai:2023cft}, making them an indispensable probe of BH nature.
	
	The study of stellar-mass particle orbits in BH spacetimes, governed by timelike geodesics, remains a cornerstone of relativistic astrophysics \cite{Kostic:2012zw,Hackmann:2008tu,Fujita:2009bp}. Notably, current research highlights the unique characteristics exhibited by bound orbits \cite{Glampedakis:2002ya,Barack:2003fp,GRAVITY:2020gka}, underscoring their importance in probing BH physics. Periodic orbits, as a critical subset of bound orbits, are especially valuable for exploring gravitational wave emission and orbital dynamics \cite{Glampedakis:2002ya,Lake:2003tr}. Following the landmark detection of gravitational waves \cite{LIGOScientific:2016aoc}, the analysis of periodic orbits has garnered substantial attention. A pivotal advancement in this domain is the classification method introduced by Levin et al. \cite{Levin:2008mq}. This approach utilizes three integers—$z$, $w$, and $v$—representing the zoom, whirl, and vertex numbers, respectively. The definition of a rational number $q = w + \frac{v}{z}$ establishes a direct mapping between mathematical classification and physical orbits, thereby achieving systematic categorization. Owing to its efficacy, this classification scheme has become a widely recognized tool in contemporary research, as demonstrated by its application in numerous studies \cite{Zhao:2024exh,Wang:2025hla,Huang:2024oli,Meng:2024cnq}.

	The paper is organized as follows. Section~\ref{section2} presents the spacetime metric for a Schwarzschild BH surrounded by a cloud of strings and embedded in perfect fluid dark matter. We derive the timelike geodesic equations and explore the dynamics of test particles, paying particular attention to the locations of the ISCOs and MBOs. In Sec.~\ref{section3}, we examine periodic orbits by characterizing their topology via the rational number $q$, and investigate how the parameters $a$ and $\alpha$ affect the energy and orbital angular momentum. The analysis of gravitational waveforms, generated by an extreme mass-ratio inspiral (EMRI) system following such periodic orbits, is presented in Sec.~\ref{section4}. Finally, we conclude in Sec.~\ref{section5} by summarizing our results and discussing their significance.
	\section{The Spacetime Metric and Timelike Geodesics of a Schwarzschild Black Hole Surrounded by a Cloud of Strings Embedded in Perfect Fluid Dark Matter}
	\label{section2}
	In this section, we introduce the geodesic orbits around a Schwarzschild BH surrounded by a cloud of strings embedded in perfect fluid dark matter. The corresponding spacetime metric is given by:
	\begin{equation}
		ds^{2}=-f(r)dt^{2}+\frac{dr^{2}}{f(r)}+r^{2}\left(d\theta^{2}+\sin^{2}\theta d \phi^{2}\right),\label{xianyuan}
	\end{equation}
	where \cite{Hamil:2024nrv}
	\begin{equation}
		f(r)=1-a-\frac{2M}{r}+\frac{\alpha}{r}\ln\frac{r}{|\alpha|}.\label{fr}
	\end{equation}
    Here, $M$ denotes the BH mass, while $a$ and $\alpha$ are the parameters associated with the cloud of strings and the perfect fluid dark matter, respectively. In the limit where $a \to 0$ and $\alpha \to 0$, the metric reduces to the standard Schwarzschild solution.
	
\begin{figure}[htbp]
		\centering
		\begin{subfigure}{0.49\textwidth}
			\includegraphics[width=3.5in, height=3.5in, keepaspectratio]{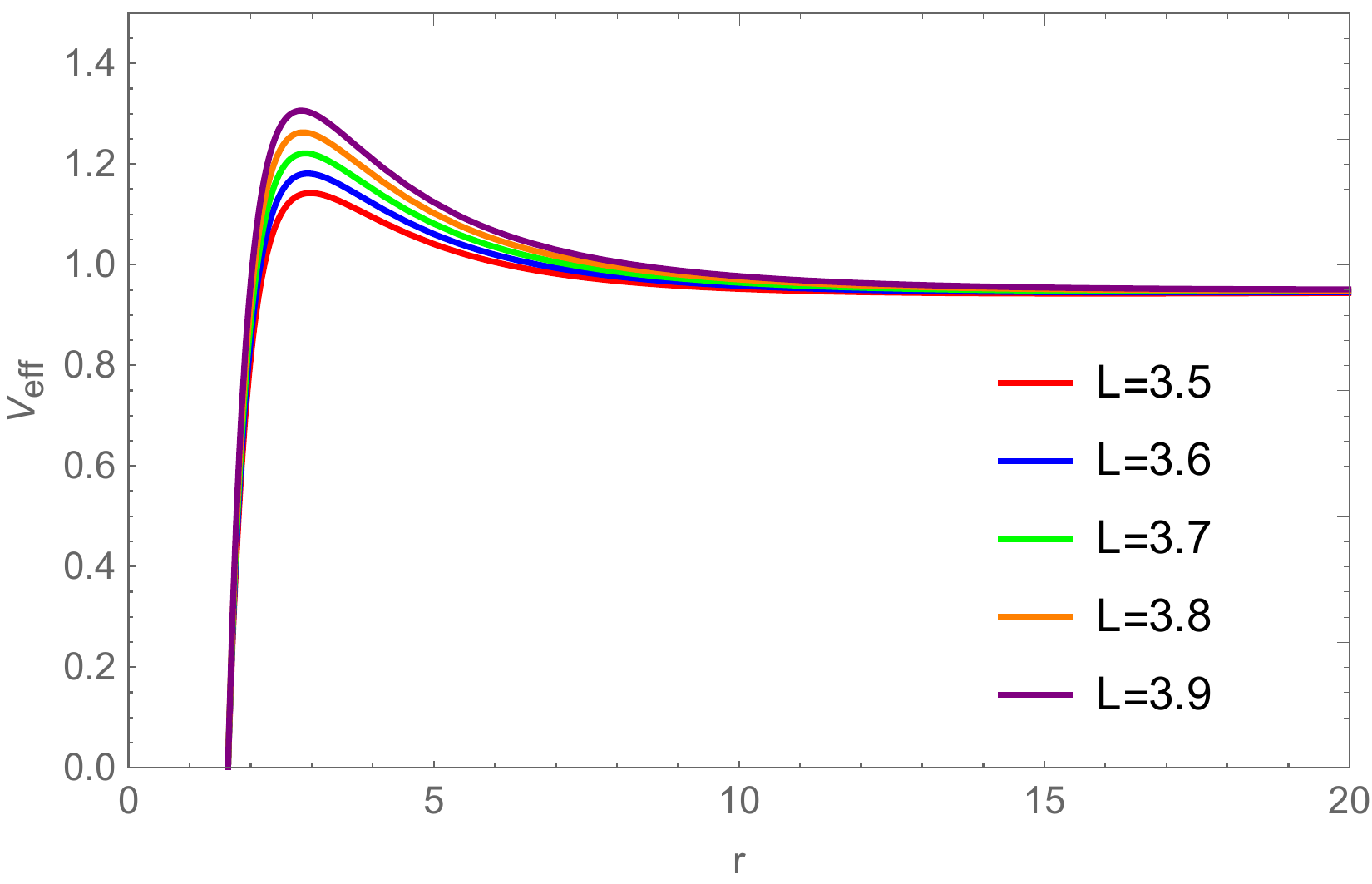}
		\end{subfigure}
		\caption{The behavior of the effective potential $V_{\text{eff}}$ as a function of the radial coordinate $r$ for varying specific angular momentum $L$. Calculations are performed for a Schwarzschild BH surrounded by a cloud of strings and embedded in perfect fluid dark matter, with $a = 0.03$ and $\alpha = 0.2$.}
		\label{veff}
	\end{figure}

	We now focus on the geodesic motion of a massive particle around a Schwarzschild BH surrounded by a cloud of strings in perfect fluid dark matter. By confining the motion to the equatorial plane ($\theta = \pi/2$), we obtain: 
	\begin{equation}
		\mathcal{L}=\frac{1}{2}g_{\mu\nu}\dot{x}^\mu\dot{x}^\nu=\varepsilon,\label{La}
	\end{equation}
	with $\varepsilon = -1$ for timelike geodesics (massive particles) and $\varepsilon = 0$ for null geodesics (massless particles). Let $\dot{x}^\mu = dx^\mu/d\tau$ denote the four-velocity tangent to the worldline, parametrized by the affine parameter $\tau$. The absence of explicit $t$ and $\phi$ terms in the Lagrangian signifies the existence of two Killing vectors, $\partial_t$ and $\partial_\phi$, which guarantee the conservation of the specific energy $E$ and specific angular momentum $L$.
	\begin{equation}
		E=f(r)\dot{t},\label{e}
	\end{equation}
	\begin{equation}
		L=r^{2}\dot{\phi},\label{l}
	\end{equation}
	For massive particles following timelike trajectories ($g_{\mu\nu}\dot{x}^\mu\dot{x}^\nu = -1$), the Lagrangian reduces to:
	\begin{equation}
		\frac{\dot{r}^{2}}{f(r)}+\frac{L^{2}}{r^{2}}-\frac{E^{2}}{f(r)}=-1,\label{la1}
	\end{equation}
	Equation (\ref{la1}) can be rewritten in the following form:
	\begin{equation}
		\dot{r}^{2}=E^{2}-V_{\text{eff}},\label{rdian2}
	\end{equation}
	where
	\begin{equation}
		V_{\text{eff}}=f(r)\left(1+\frac{L^{2}}{r^{2}}\right).\label{Veff}
	\end{equation}
	Here, $V_{\text{eff}}(r)$ represents the radial effective potential determining the particle's radial evolution. Figure \ref{veff} reveals a pronounced dependence of the effective potential on the particle's angular momentum $L$. As depicted in Fig. \ref{veff}, increasing the angular momentum $L$ leads to a higher effective potential barrier. 
	
\begin{figure*}[htbp]
		\centering
		\begin{subfigure}{0.38\textwidth}
			\includegraphics[width=\linewidth]{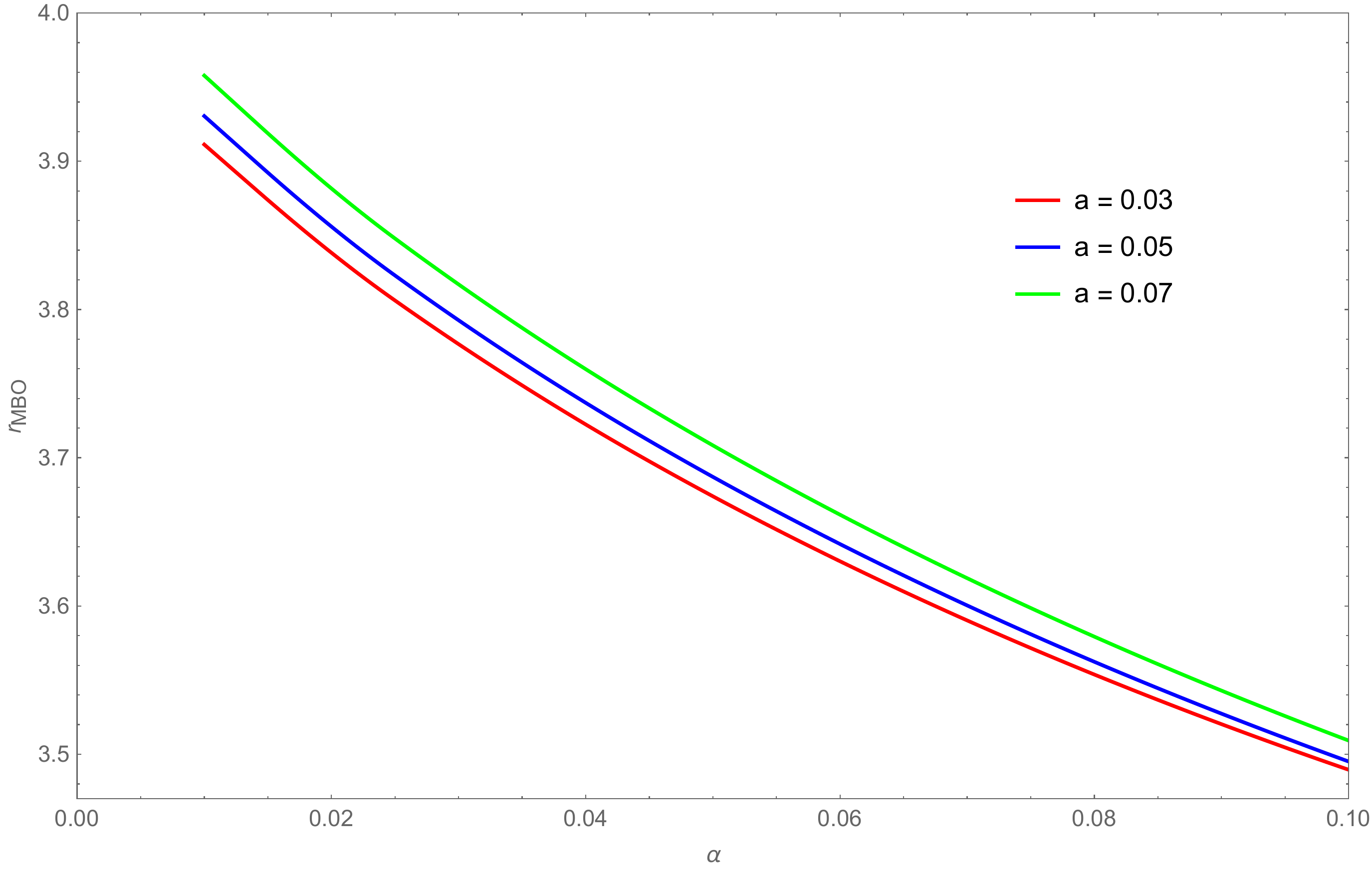}
		\end{subfigure}%
		\hspace{0.13\textwidth}% 
		\begin{subfigure}{0.38\textwidth}
			\includegraphics[width=\linewidth]{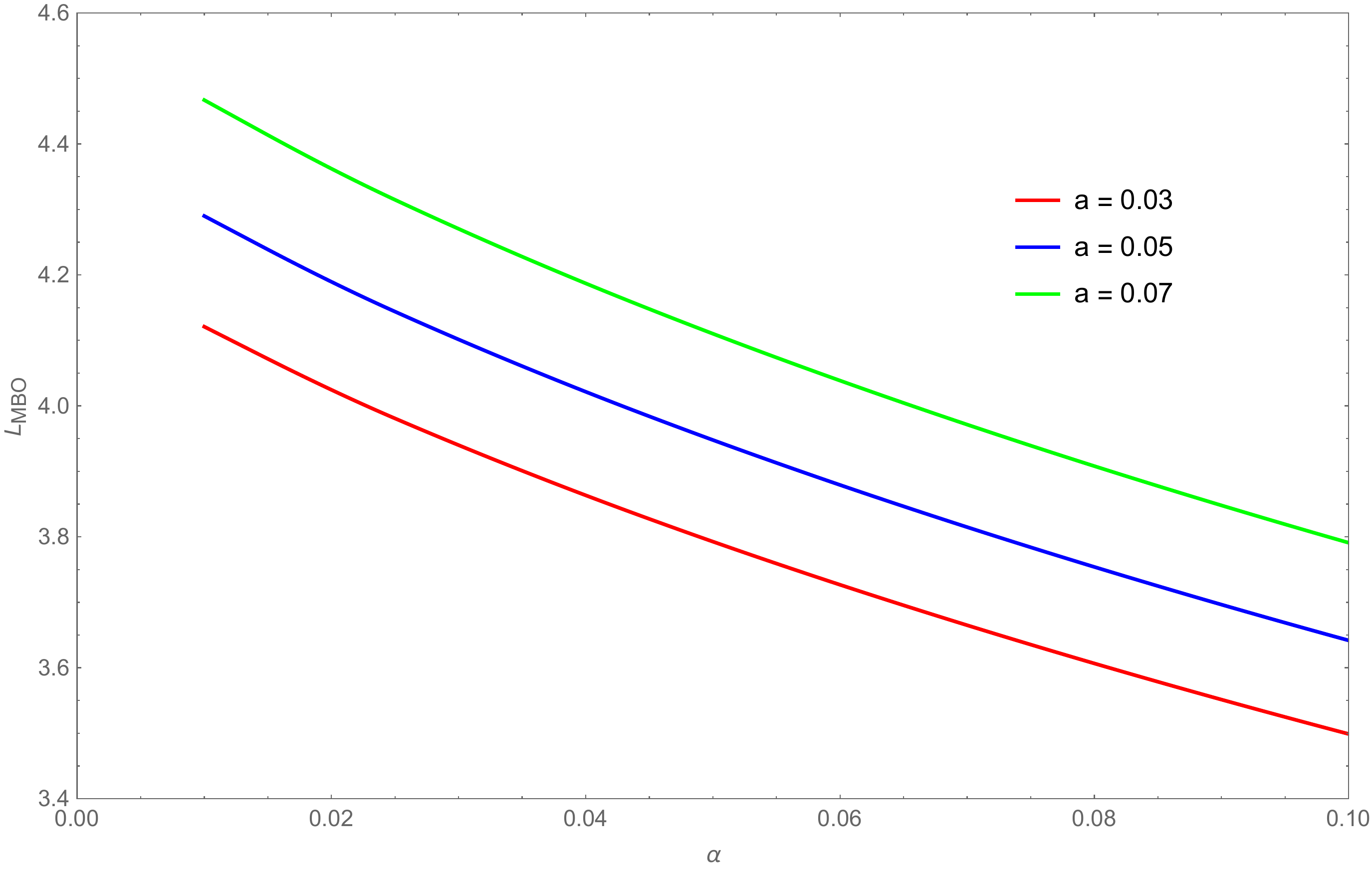}
		\end{subfigure}
		\caption{Variation of the MBO radius $r_{\text{MBO}}$ and angular momentum $L_{\text{MBO}}$ with respect to the parameter $\alpha$.}
		\label{rlmbo}
	\end{figure*}

	While periodic orbits are a particular type of bound orbit, the general condition for a particle to remain in a bound state requires its energy and orbital angular momentum to obey:
	\begin{equation}
		L_{\mathrm{ISCO}} \leq L \quad \text{and} \quad E_{\mathrm{ISCO}} \leq E \leq E_{\mathrm{MBO}} = 1. \label{range}
	\end{equation}
	In these expressions, $E_{\text{ISCO}}$ ($L_{\text{ISCO}}$) and $E_{\text{MBO}}$ correspond to the energy (angular momentum) of the ISCO and the energy of the MBO, respectively. 
	Since the MBO and ISCO define the boundaries of bound motion, examining their dependence on the string cloud and perfect fluid dark matter parameters provides key insights into the behavior of periodic orbits. 
	Specifically, the MBO is determined by the condition:
	\begin{equation}
		V_{\text{eff}} = 1,\ \frac{dV_{\text{eff}}}{dr} = 0.\label{mbo}
	\end{equation}
	Building on the preceding equations, we numerically investigate the dependence of the MBO's radius and orbital angular momentum on the parameters characterizing the string cloud and perfect fluid dark matter. In Fig. \ref{rlmbo}, we plot the dependence of the MBO radius and orbital angular momentum on the perfect fluid dark matter parameter, considering different strengths of the string cloud. It is evident from the plots that both the radius and angular momentum exhibit an upward trend as the string cloud parameter increases.
	
\begin{figure*}[htbp]
		\centering
		\begin{subfigure}{0.32\textwidth}
			\centering
			\includegraphics[width=\linewidth]{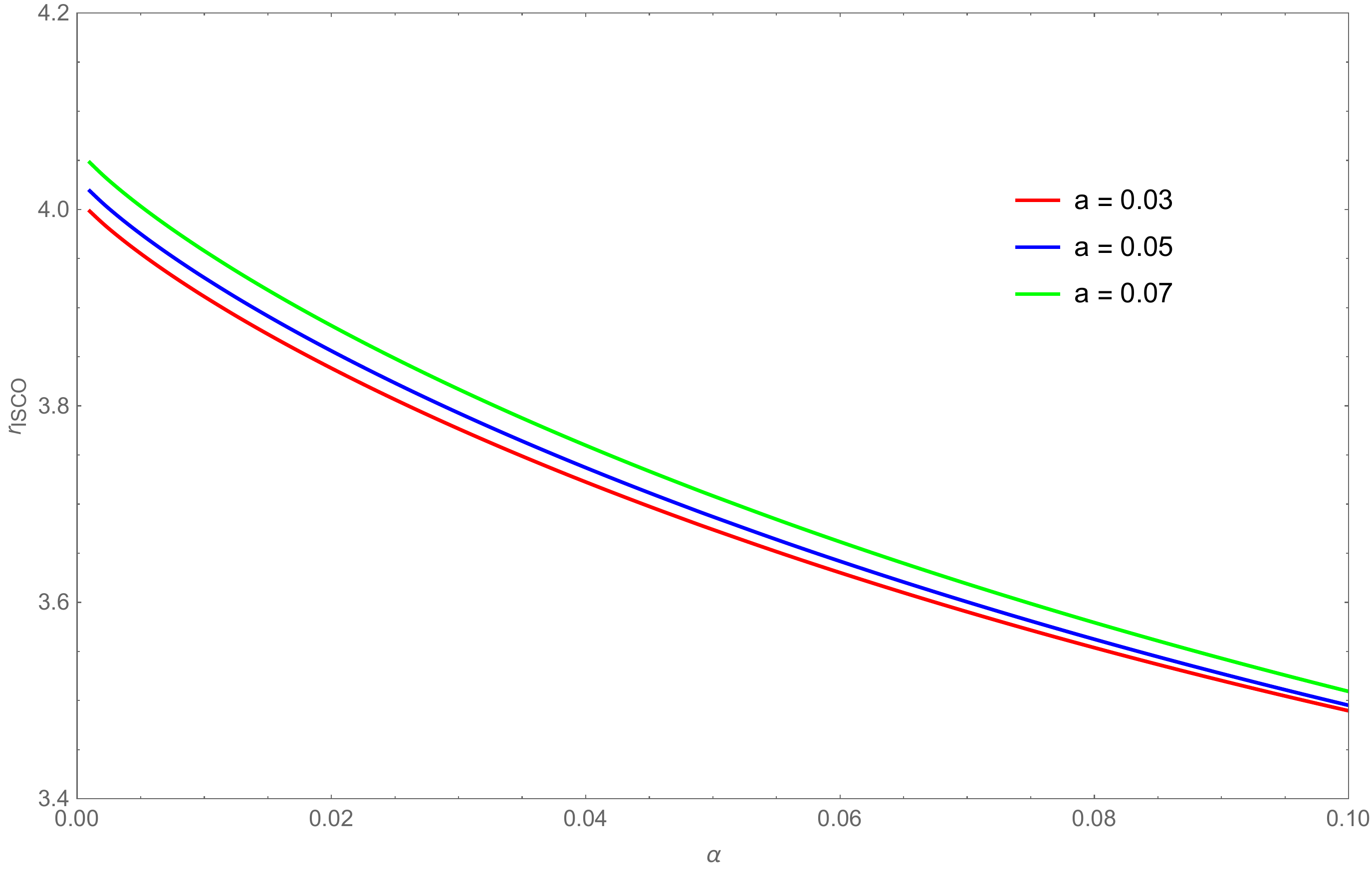}
		\end{subfigure}%
		\hfill
		\begin{subfigure}{0.32\textwidth}
			\centering
			\includegraphics[width=\linewidth]{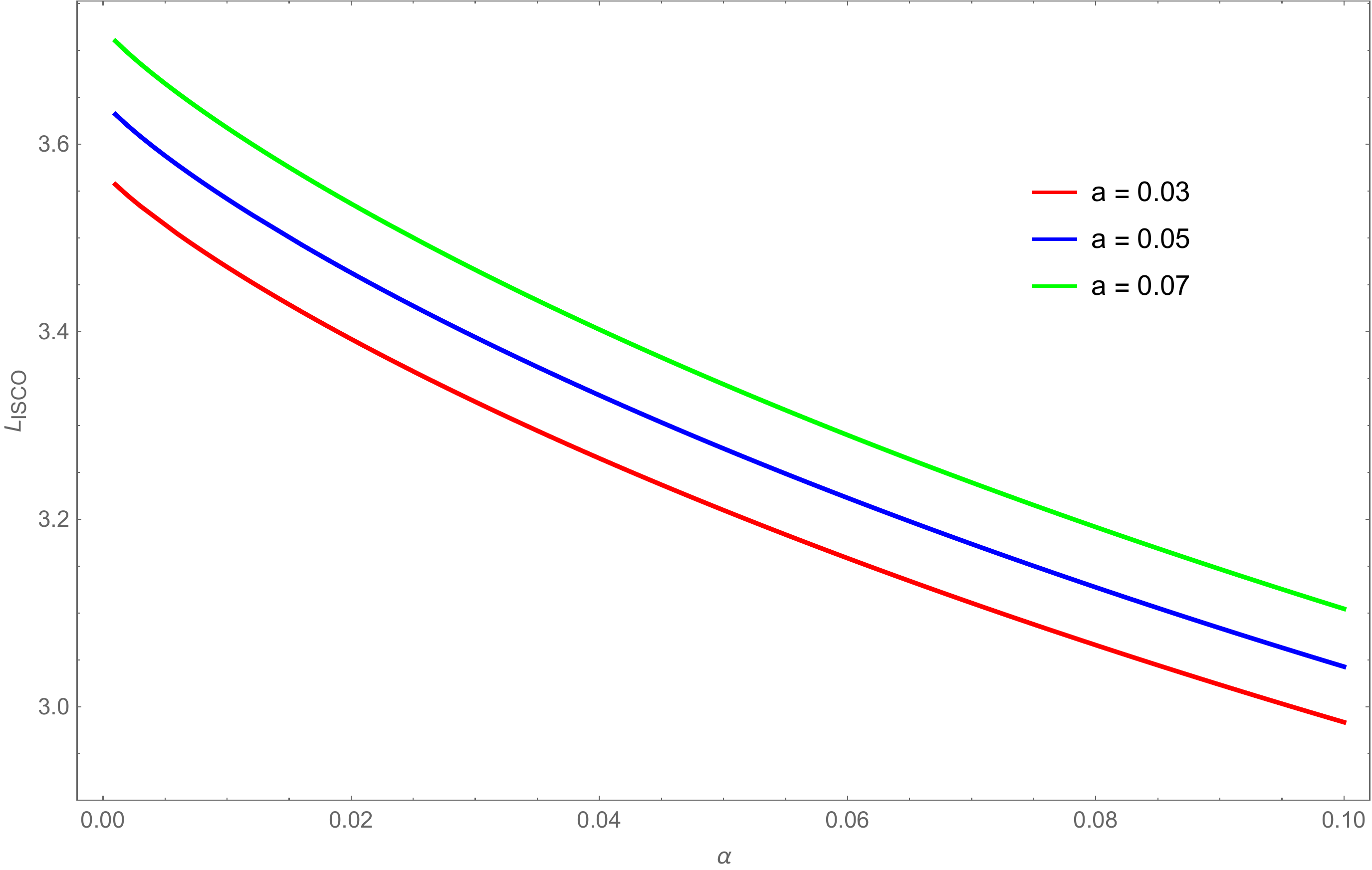}
		\end{subfigure}%
		\hfill
		\begin{subfigure}{0.32\textwidth}
			\centering
			\includegraphics[width=\linewidth]{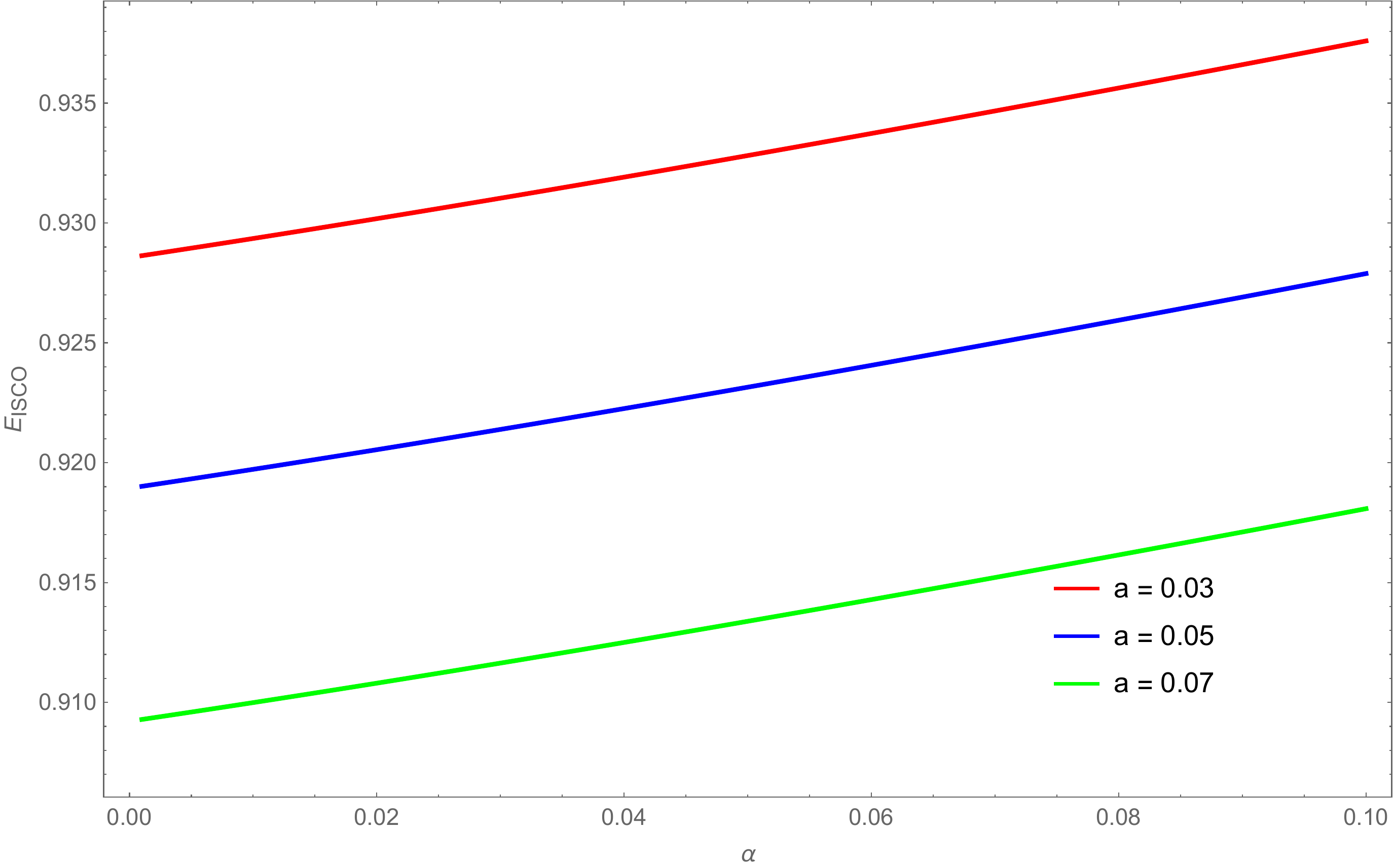}
		\end{subfigure}
		
		\caption{Variation of the ISCO characteristics ($r_{\text{ISCO}}$, $L_{\text{ISCO}}$, $E_{\text{ISCO}}$) with respect to the parameter $\alpha$.}
		\label{rleisco}
	\end{figure*}

\begin{figure*}[htbp]
		\centering
		\begin{subfigure}{0.38\textwidth}
			\includegraphics[width=\linewidth]{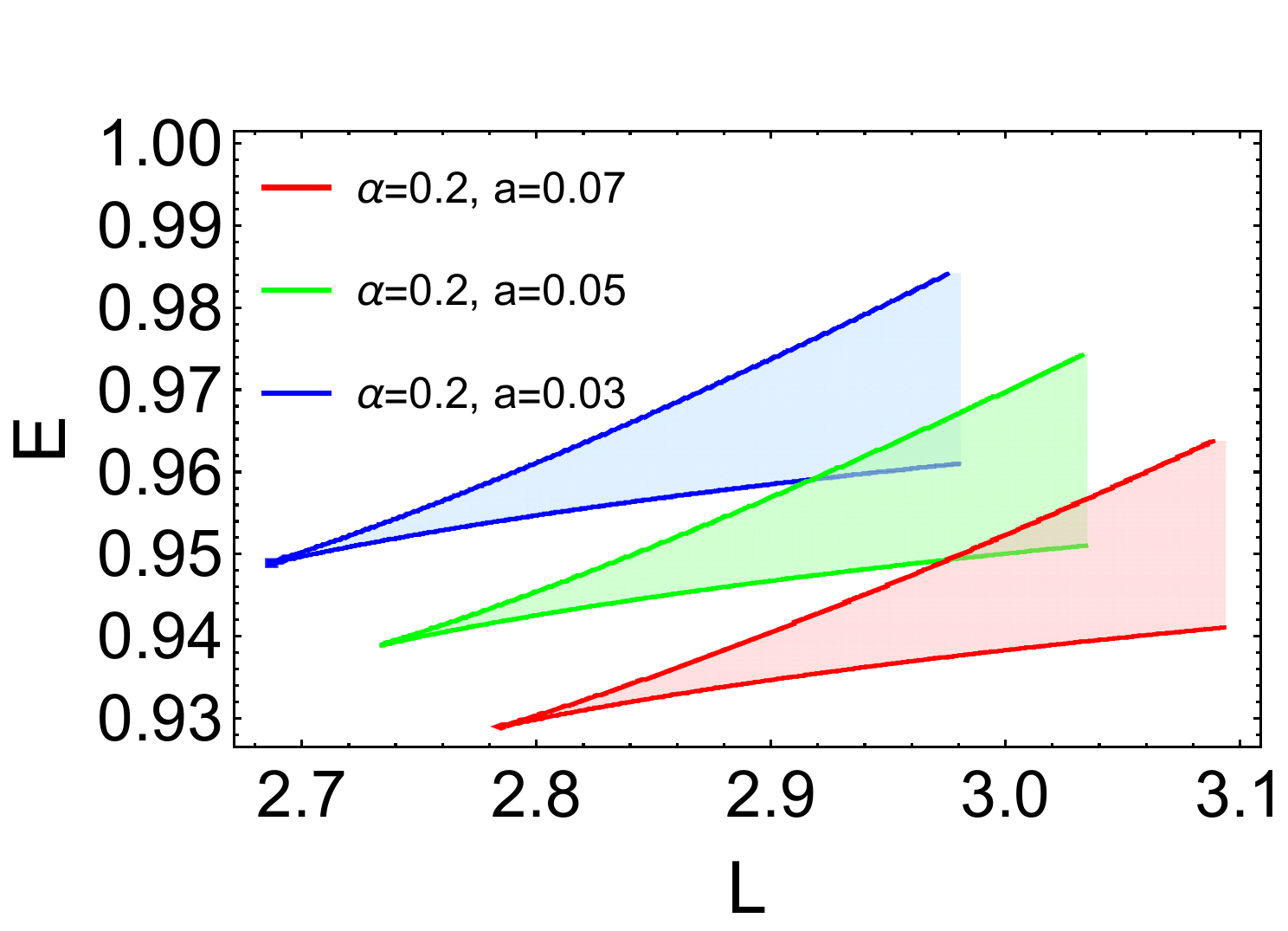}
		\end{subfigure}%
		\hspace{0.13\textwidth}% 控制间距
		\begin{subfigure}{0.38\textwidth}
			\includegraphics[width=\linewidth]{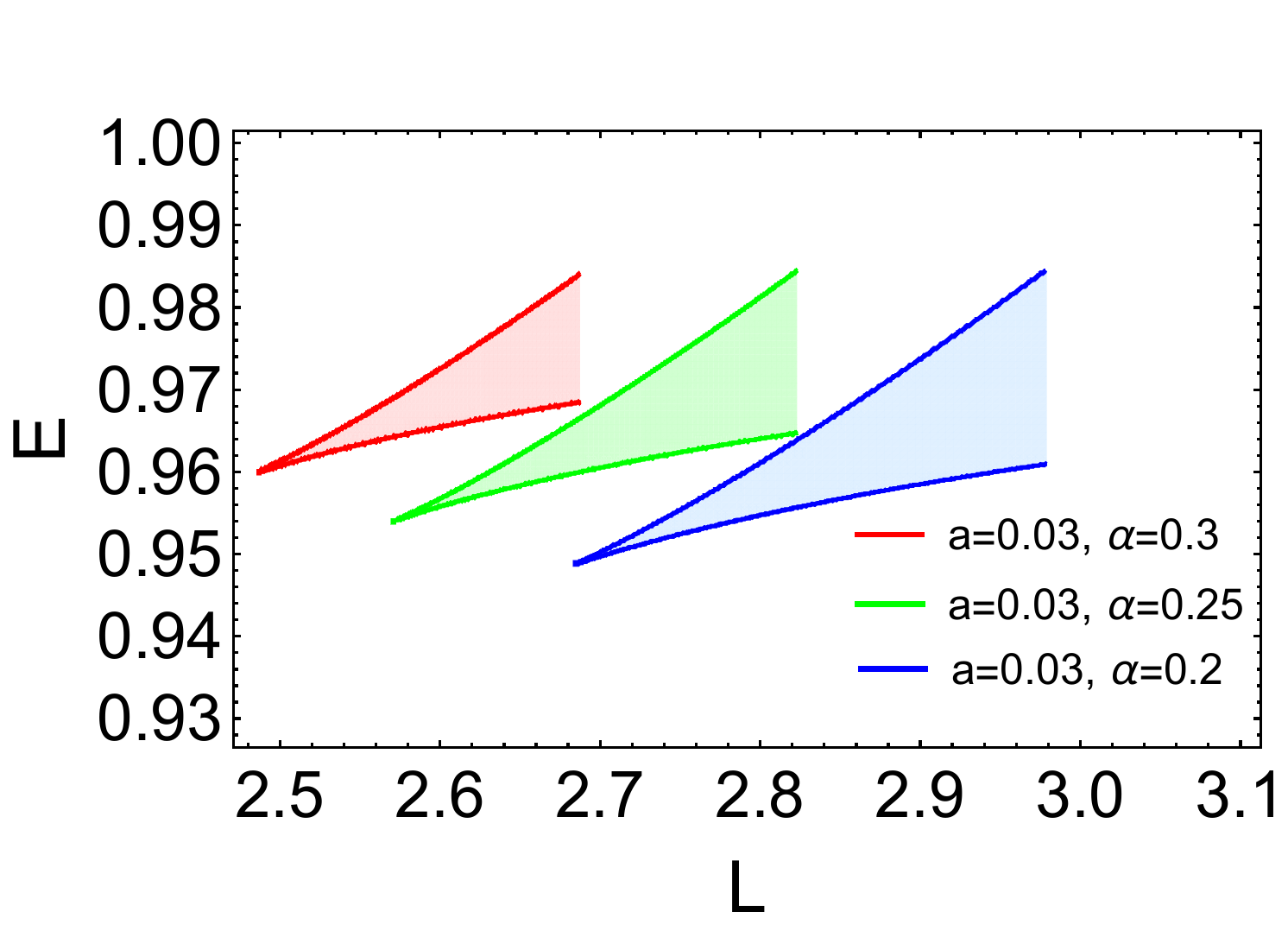}
		\end{subfigure}
		
		\caption{The allowed parameter space of energy $E$ and orbital angular momentum $L$ for bound orbits around a Schwarzschild BH surrounded by a cloud of strings and embedded in perfect fluid dark matter, for different values of the spacetime parameters. Left: $\alpha = 0.2$; Right: $a = 0.03$.}
		\label{el}
	\end{figure*}

	Having analyzed the properties of the MBO, we now turn our attention to the ISCO. Mathematically, the location of this critical orbit is governed by the following conditions:
	\begin{equation}
		\dot{r} = 0,\quad \frac{dV_{\text{eff}}}{dr} = 0 \quad \text{and} \quad \frac{d^{2}V_{\text{eff}}}{dr^{2}} = 0.\label{isco0}
	\end{equation}
	Due to the spherical symmetry of the metric, the ISCO radius ($r_{\text{ISCO}}$) can be determined by solving Eq. (\ref{isco0}). The corresponding angular momentum ($L_{\text{ISCO}}$) and energy ($E_{\text{ISCO}}$) are then given by:
	\begin{equation}
		r_{\text{ISCO}} =\frac{3f(r_{\text{ISCO}})f^{\prime}(r_{\text{ISCO}})}{2[f^{\prime}(r_{\text{ISCO}})]^{2}-f(r_{\text{ISCO}})f^{\prime\prime}(r_{\text{ISCO}})},\label{Risco}
	\end{equation}
	\begin{equation}
		L_{\text{ISCO}} =\sqrt{\frac{r_{\text{ISCO}}^{3}f^{\prime}(r_{\text{ISCO}})}{2f(r_{\text{ISCO}})-r_{\text{ISCO}}f^{\prime}(r_{\text{ISCO}})}},\label{Lisco}
	\end{equation}
	\begin{equation}
		E_{\text{ISCO}} =\frac{f(r_{\text{ISCO}})}{\sqrt{f(r_{\text{ISCO}})-\frac{1}{2}r_{\text{ISCO}}f^{\prime}(r_{\text{ISCO}})}},\label{Eisco}
	\end{equation}
	These expressions explicitly depend on the dark matter parameter and the string cloud parameter, reflecting the modification of the spacetime geometry. Figure \ref{rleisco} illustrates the dependence of the ISCO radius ($r_{\text{ISCO}}$), specific angular momentum ($L_{\text{ISCO}}$), and specific energy ($E_{\text{ISCO}}$) on the perfect fluid dark matter parameter $\alpha$ for various values of the string cloud parameter $a$. As shown in the figure, both $r_{\text{ISCO}}$ and $L_{\text{ISCO}}$ increase monotonically with $a$, whereas they decrease as $\alpha$ increases. In contrast, $E_{\text{ISCO}}$ exhibits an inverse behavior: it diminishes with increasing $a$ but rises with larger $\alpha$. These contrasting trends can be physically interpreted as the competition between the two exotic matter fields modifying the effective gravitational potential. The presence of the string cloud ($a$) appears to weaken the effective attraction at the orbital scale, pushing the marginally stable orbit outward and requiring higher angular momentum for equilibrium. Conversely, the perfect fluid dark matter component ($\alpha$) seems to deepen the potential well or enhance the central attraction, permitting stable orbits to persist closer to the BH with reduced angular momentum, though at a higher total energy state.
	Figure \ref{el} delineates the allowed parameter space for energy $E$ and orbital angular momentum $L$ of bound orbits, derived from Eq. (\ref{rdian2}). It is observed that, for a fixed dark matter parameter $\alpha$, an increase in the string cloud parameter $a$ shifts the allowed region notably toward the lower-right direction. In contrast, for a fixed $a$, an increase in $\alpha$ moves the allowed region markedly toward the upper-left direction. Physically, the string cloud parameter $a$ effectively weakens the gravitational binding near the BH, permitting bound orbits with lower energy and higher angular momentum. On the other hand, the perfect fluid dark matter parameter $\alpha$ enhances the central gravitational attraction, thereby requiring higher energy and lower angular momentum to maintain bound motion.
	\section{Periodic orbits}
	\label{section3}
\begin{figure*}[htbp]
		\centering
		\begin{subfigure}{0.45\textwidth}
			\includegraphics[width=3in, height=5.5in, keepaspectratio]{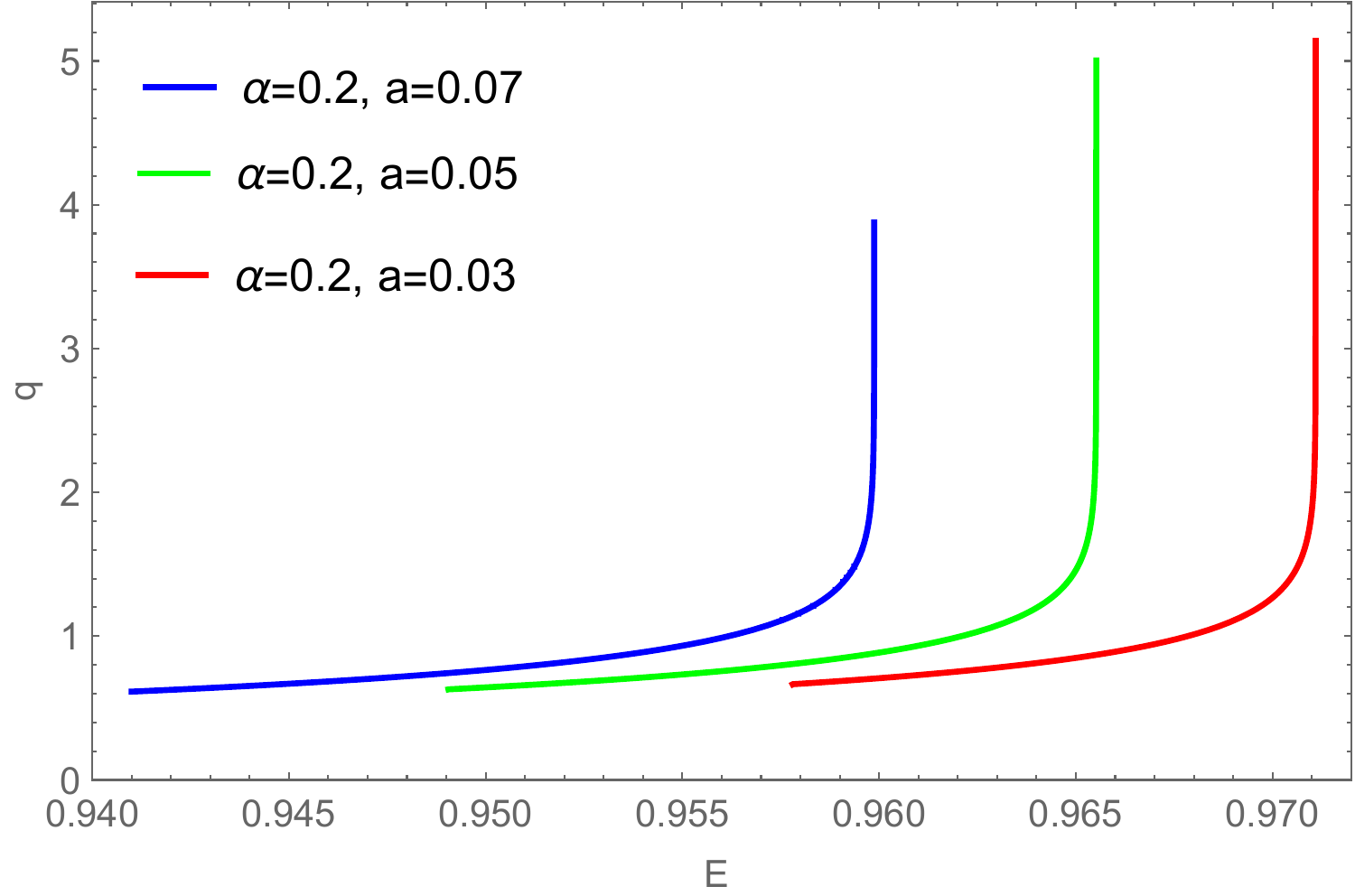}
		\end{subfigure}
		\hfill
		\begin{subfigure}{0.45\textwidth}
			\includegraphics[width=3in, height=5.5in,keepaspectratio]{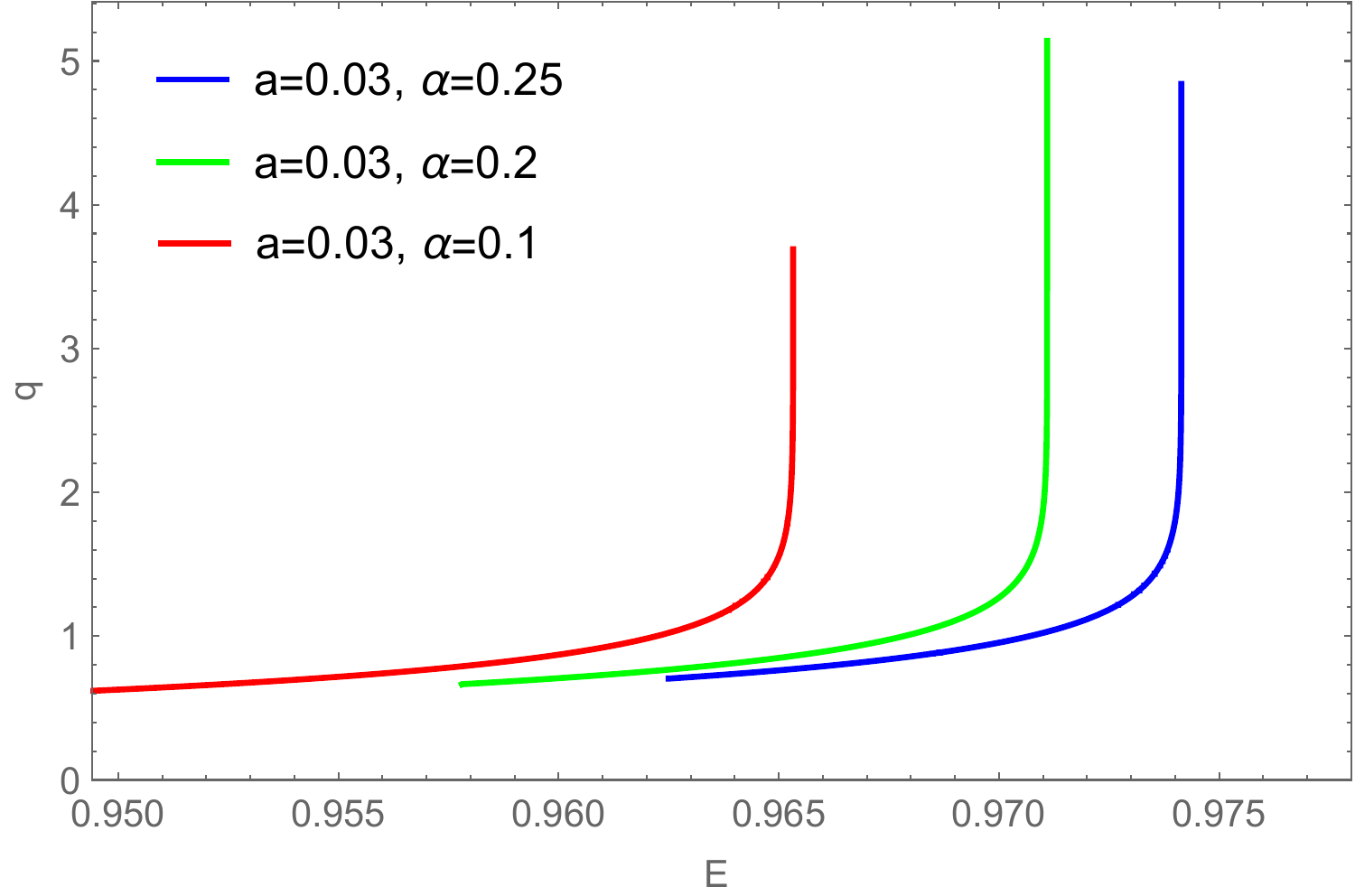}
		\end{subfigure}
		\begin{subfigure}{0.45\textwidth}
			\includegraphics[width=3in, height=5.5in, keepaspectratio]{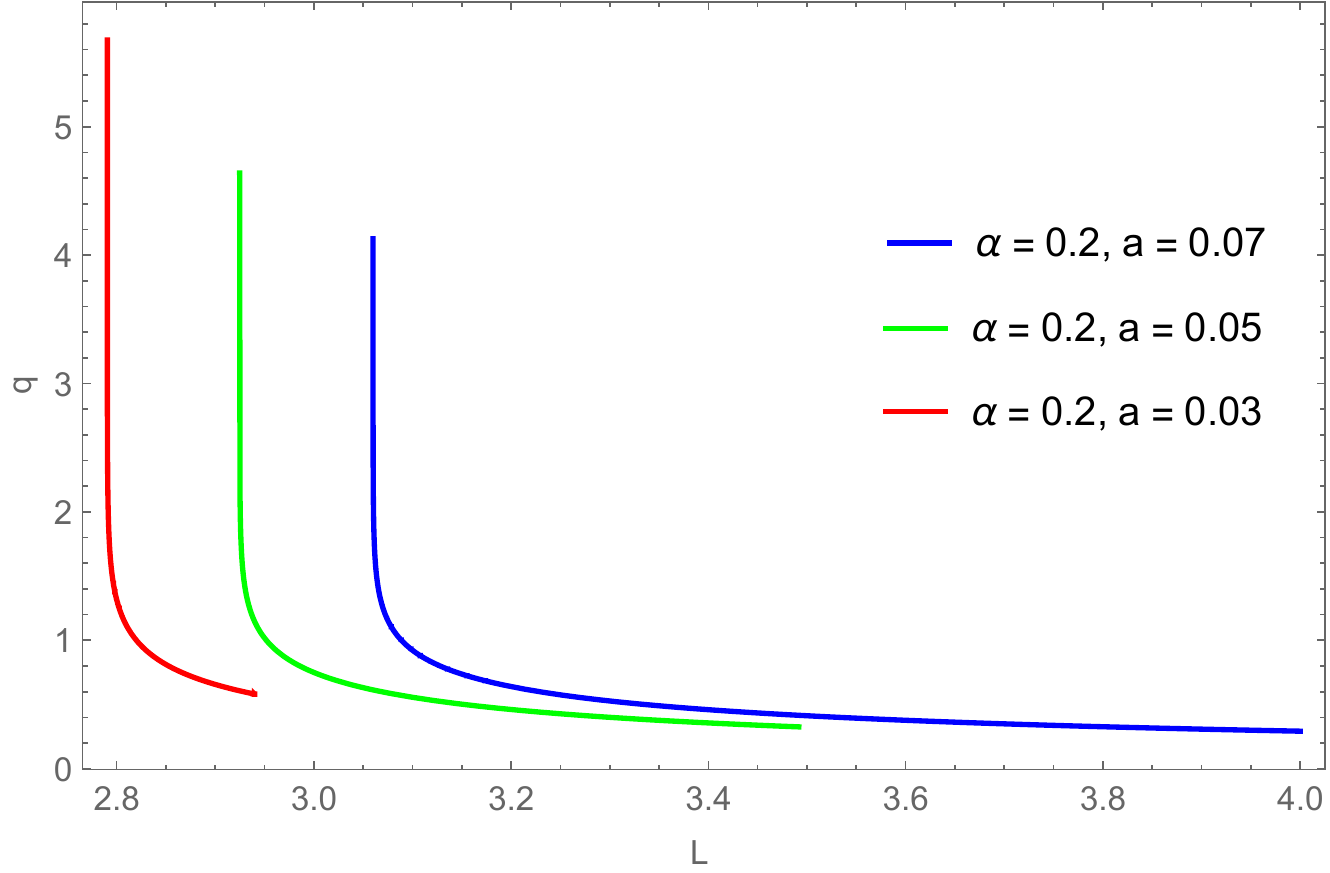}
		\end{subfigure}
		\hfill
		\begin{subfigure}{0.45\textwidth}
			\includegraphics[width=3in, height=5.5in, keepaspectratio]{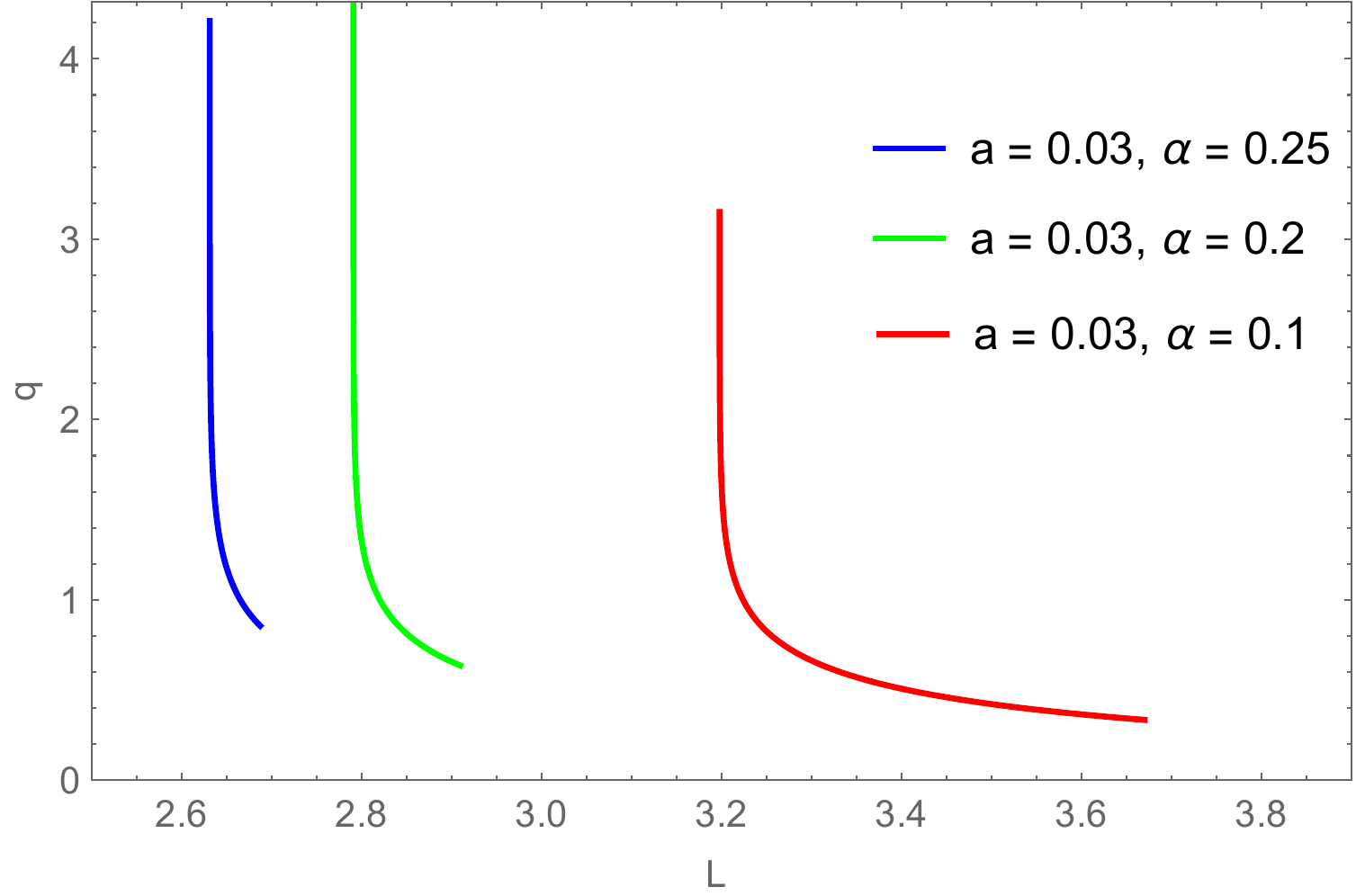}
		\end{subfigure}
		\caption{Top panel: Dependence of the rational number $q$ on the energy $E$ for periodic orbits around a Schwarzschild BH with a cloud of strings embedded in perfect fluid dark matter. The left panel shows the results for varying the string cloud parameter $a$ (with fixed dark matter parameter $\alpha = 0.2$), while the right panel shows the dependence on $\alpha$ (with fixed $a = 0.03$). In both cases, the orbital angular momentum is set to $L = \frac{1}{2}(L_{\mathrm{MBO}} + L_{\mathrm{ISCO}})$. Bottom panel: The rational number $q$ as a function of the orbital angular momentum $L$, with the particle energy fixed at $E = 0.96$. Different curves correspond to distinct combinations of the parameters $\alpha$ and $a$.}
		\label{qel}
	\end{figure*}

	To investigate the properties of periodic orbits, we utilize the taxonomy proposed in Ref.~\cite{Levin:2008mq}. In this approach, each orbit is labeled via a rational number $q$ constructed from three integers $(z, w, v)$ according to:
	\begin{equation}
		q=\frac{\omega_{\phi}}{\omega_{r}}-1=w+\frac{v}{z},\label{qwvz}  
	\end{equation}
	Here, $\omega_{\phi}$ and $\omega_{r}$ denote the azimuthal and radial orbital frequencies, while $z$, $w$ and $v$ correspond to the zoom, whirl, and vertex numbers of the orbit, respectively. By combining the equations of motion for the particle, Eqs. (\ref{l}) and (\ref{rdian2}), Eq. (\ref{qwvz}) can be rewritten as:
	\begin{equation}
		q=\frac{1}{\pi}\int_{r_{1}}^{r_{2}}\frac{\dot{\phi}}{\dot{r}}-1=\frac{1}{\pi}\int_{r_{1}}^{r_{2}}\frac{1}{r^{2}\sqrt{E^{2}-V_{\text{eff}}}}dr-1,\label{qint}  
	\end{equation}
	where $r_1$ and $r_2$ correspond to the periapsis and apoapsis radii, respectively. We investigate the behavior of the rational number $q$ under two scenarios: (i) fixing the angular momentum at $L = (L_{\mathrm{MBO}} + L_{\mathrm{ISCO}})/2$ while varying the energy $E$, and (ii) fixing the energy at $E = 0.96$ while varying $L$. In both cases, we consider different values of the parameters $a$ and $\alpha$, as illustrated in Fig. \ref{qel}. From the top panel, we observe a steady increase in $q$ with increasing $E$, followed by a steep ascent as $E$ nears its upper bound. An increase in the parameter $a$ (holding $\alpha$ constant) systematically shifts the profiles to lower energies, while an increase in $\alpha$ (holding $a$ constant) systematically shifts them to higher energies. The bottom panel illustrates the variation of $q$ with respect to $L$, where $q$ is found to decrease gradually with increasing angular momentum.
	
\begin{figure*}[htbp]
		\centering
		
		\begin{subfigure}{0.3\textwidth}
			\centering
			\includegraphics[width=\linewidth, keepaspectratio]{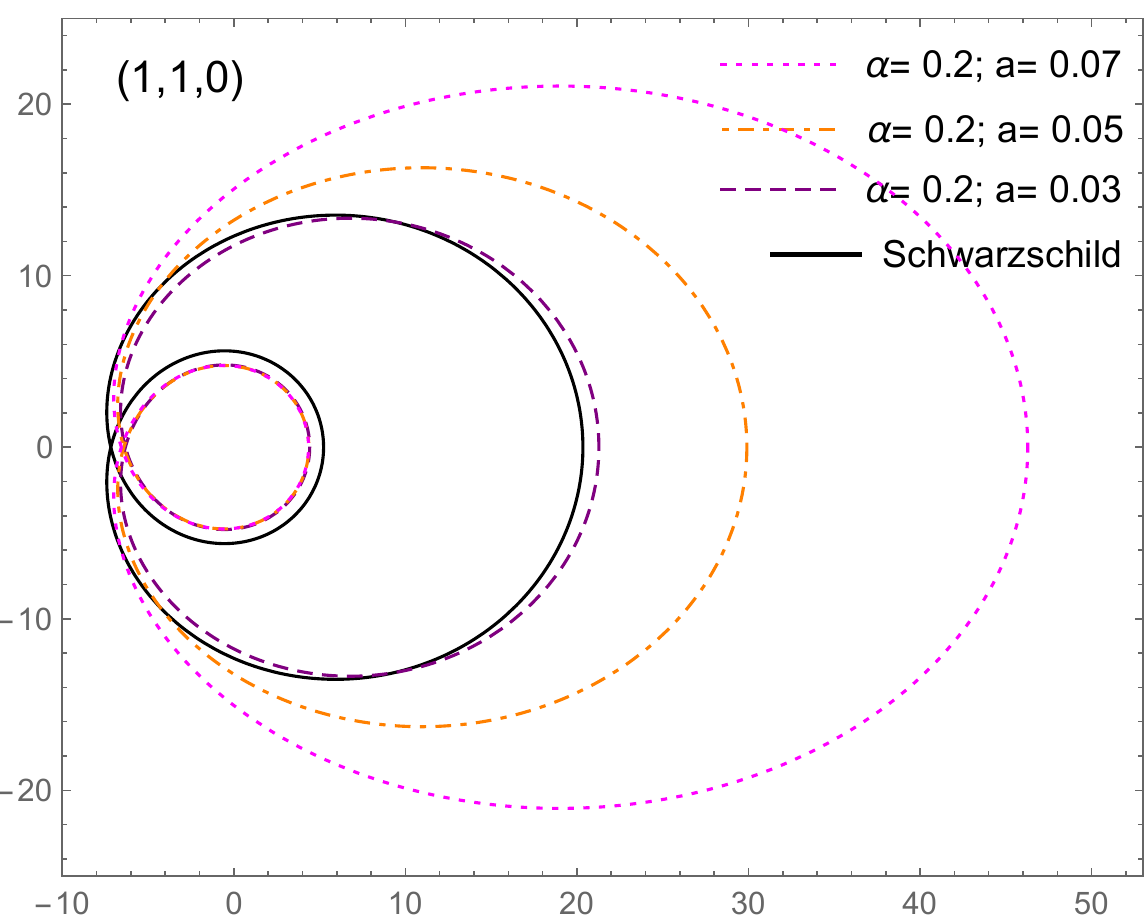}
		\end{subfigure}
		\hfill
		\begin{subfigure}{0.3\textwidth}
			\centering
			\includegraphics[width=\linewidth, keepaspectratio]{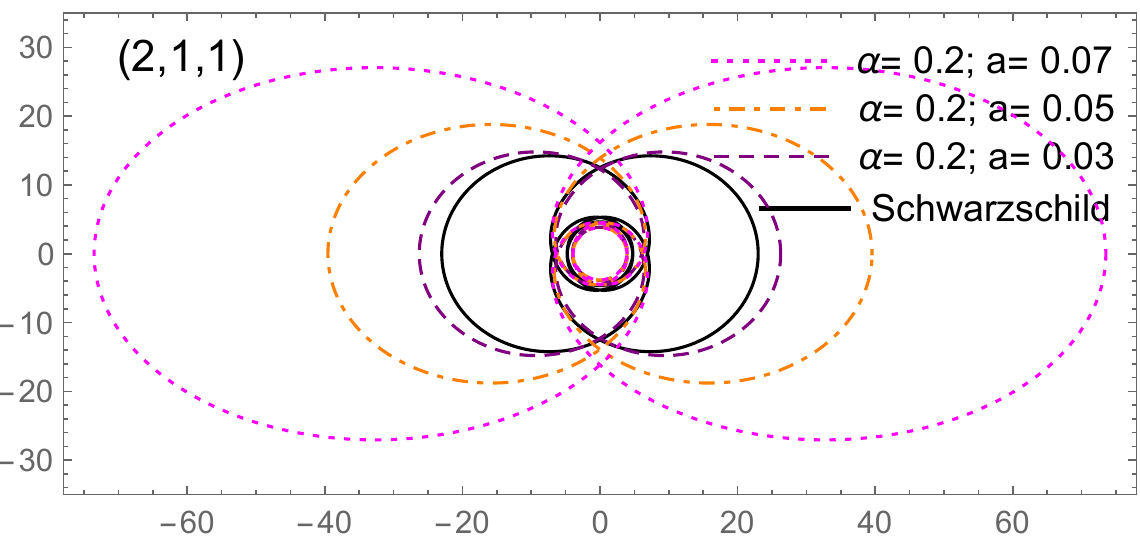}
		\end{subfigure}
		\hfill
		\begin{subfigure}{0.3\textwidth}
			\centering
			\includegraphics[width=\linewidth, keepaspectratio]{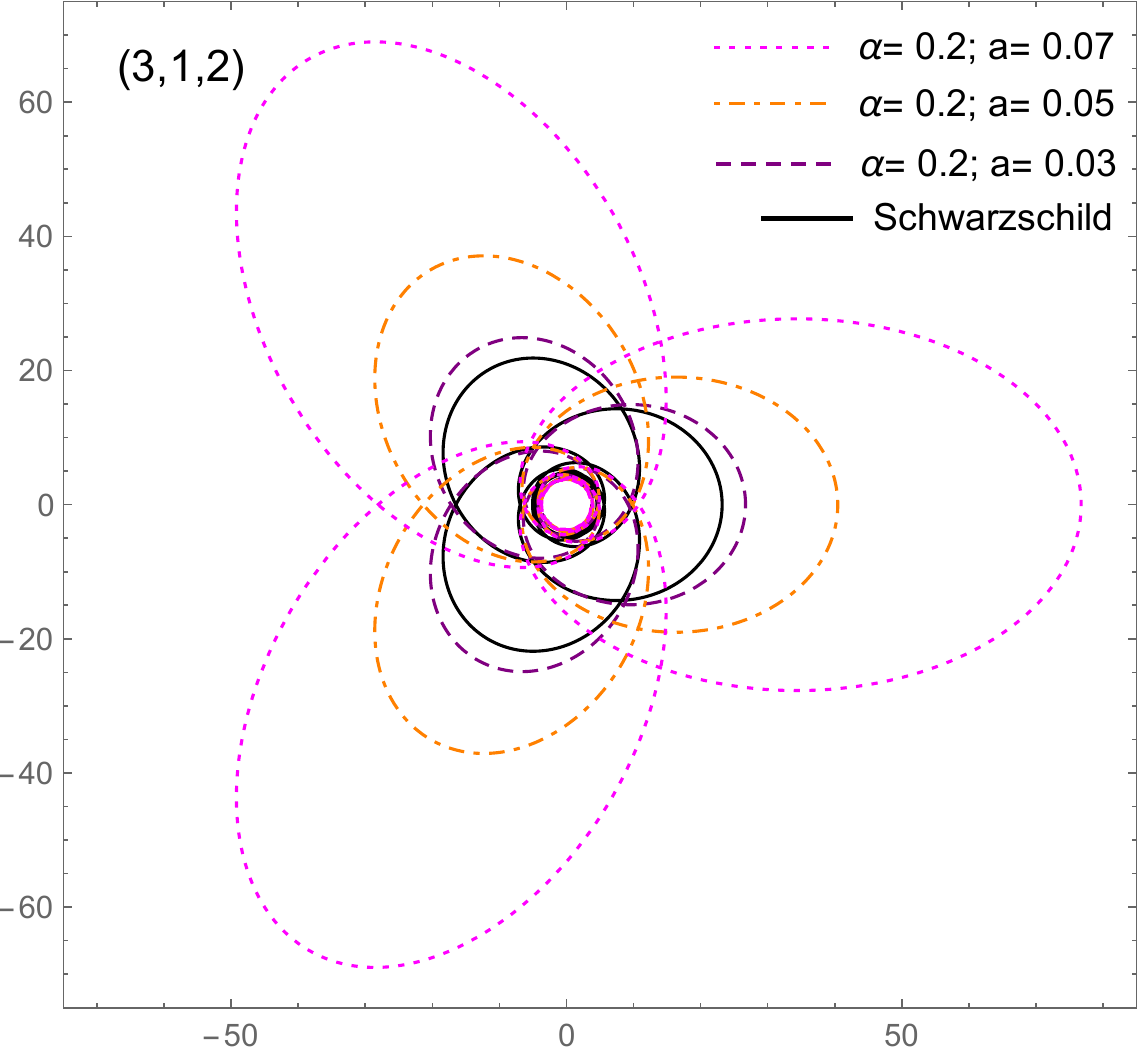}
		\end{subfigure}
		
		\vspace{0.5cm}

		\begin{subfigure}{0.3\textwidth}
			\centering
			\includegraphics[width=\linewidth, keepaspectratio]{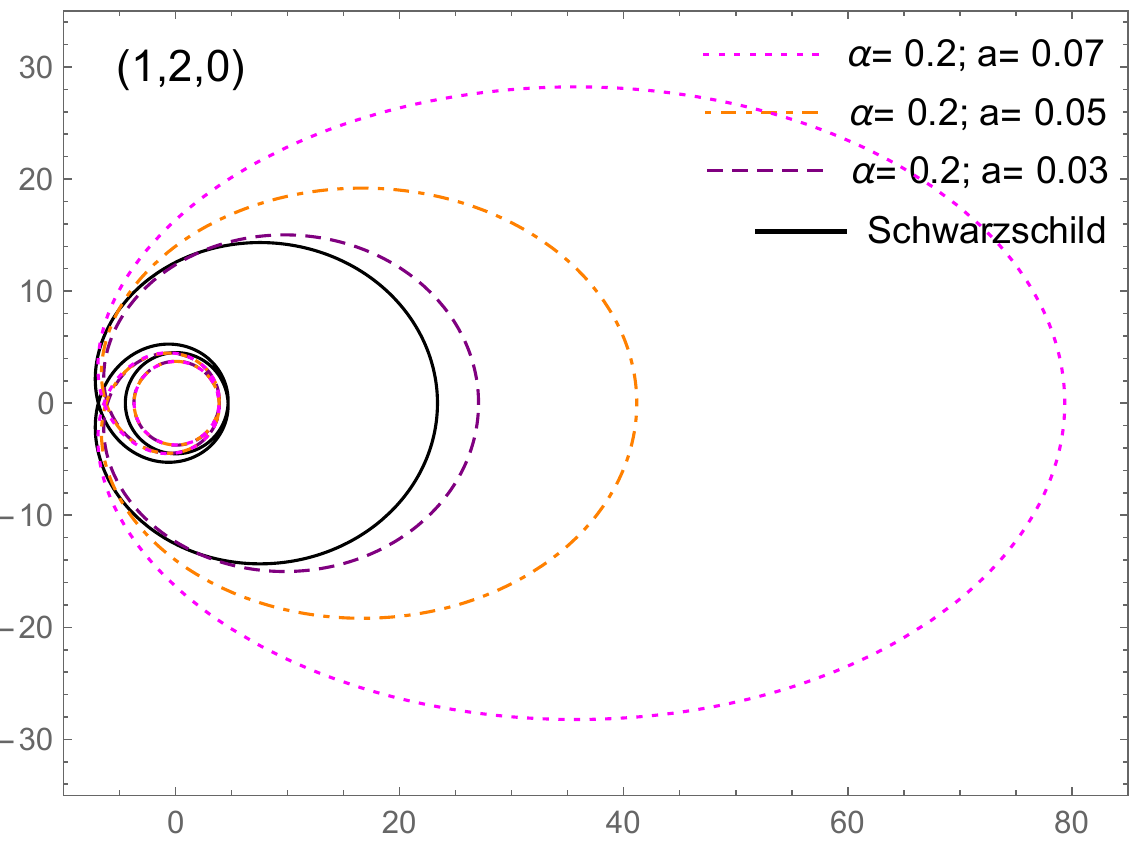}
		\end{subfigure}
		\hfill
		\begin{subfigure}{0.3\textwidth}
			\centering
			\includegraphics[width=\linewidth, keepaspectratio]{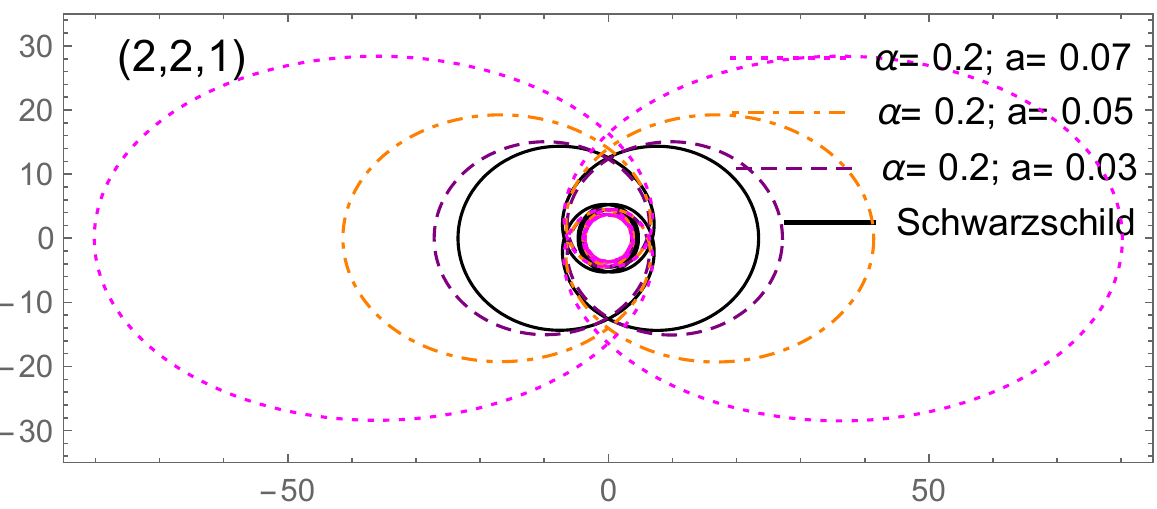}
		\end{subfigure}
		\hfill
		\begin{subfigure}{0.3\textwidth}
			\centering
			\includegraphics[width=\linewidth, keepaspectratio]{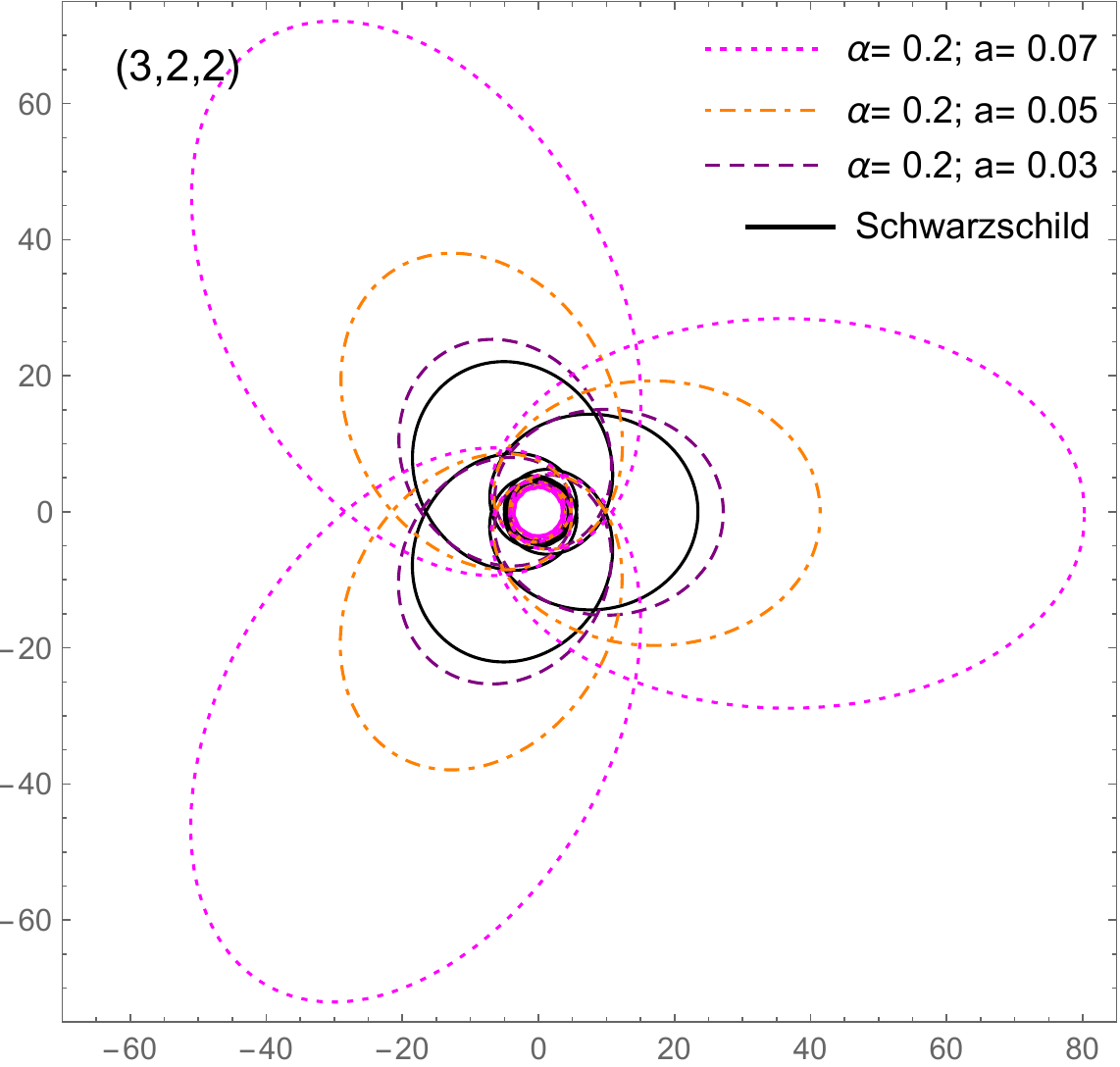}
		\end{subfigure}
		
		\caption{Periodic orbits labeled by $(z, w, v)$ around a Schwarzschild BH with a cloud of strings embedded in perfect fluid dark matter, shown for various values of the parameter $a$ with fixed $L = \frac{1}{2}(L_{\mathrm{MBO}} + L_{\mathrm{ISCO}})$ and $\alpha = 0.2$. The black curves correspond to the pure Schwarzschild case.}
		\label{orbitL}
	\end{figure*}

\begin{figure*}[htbp]
		\centering
		
		\begin{subfigure}{0.3\textwidth}
			\centering
			\includegraphics[width=\linewidth, keepaspectratio]{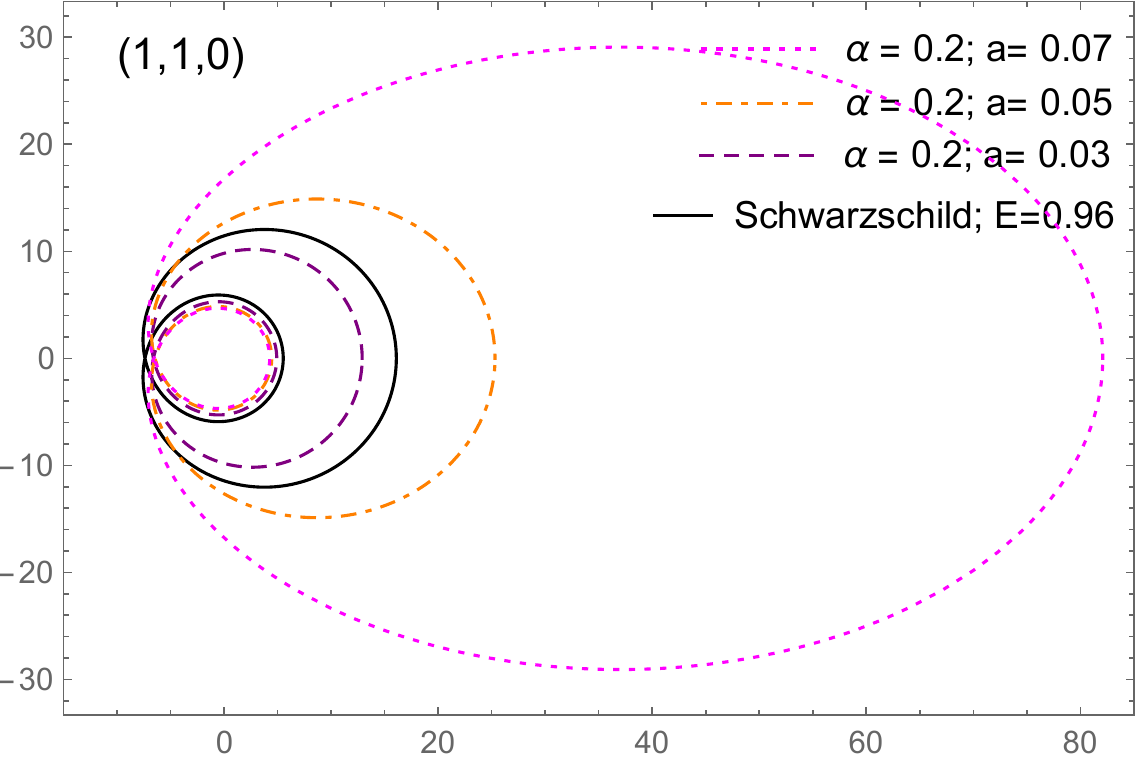}
		\end{subfigure}
		\hfill
		\begin{subfigure}{0.3\textwidth}
			\centering
			\includegraphics[width=\linewidth, keepaspectratio]{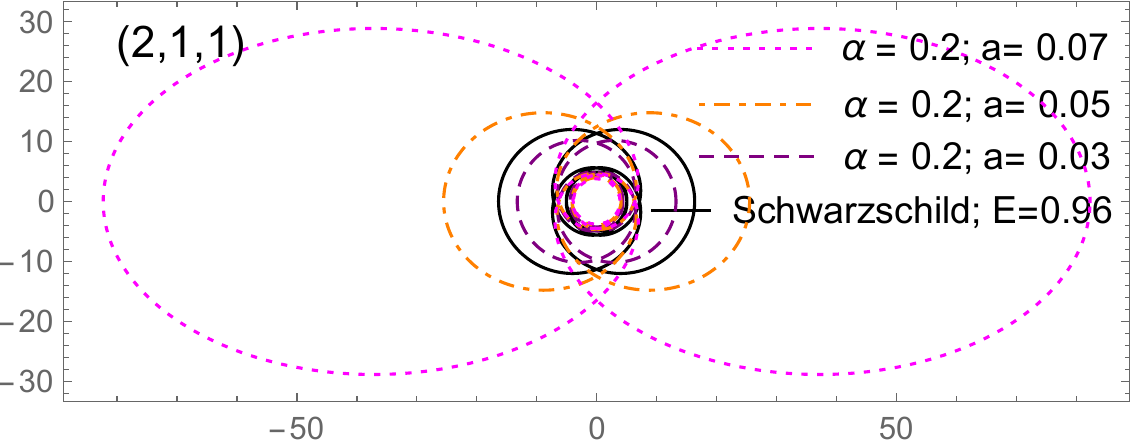}
		\end{subfigure}
		\hfill
		\begin{subfigure}{0.3\textwidth}
			\centering
			\includegraphics[width=\linewidth, keepaspectratio]{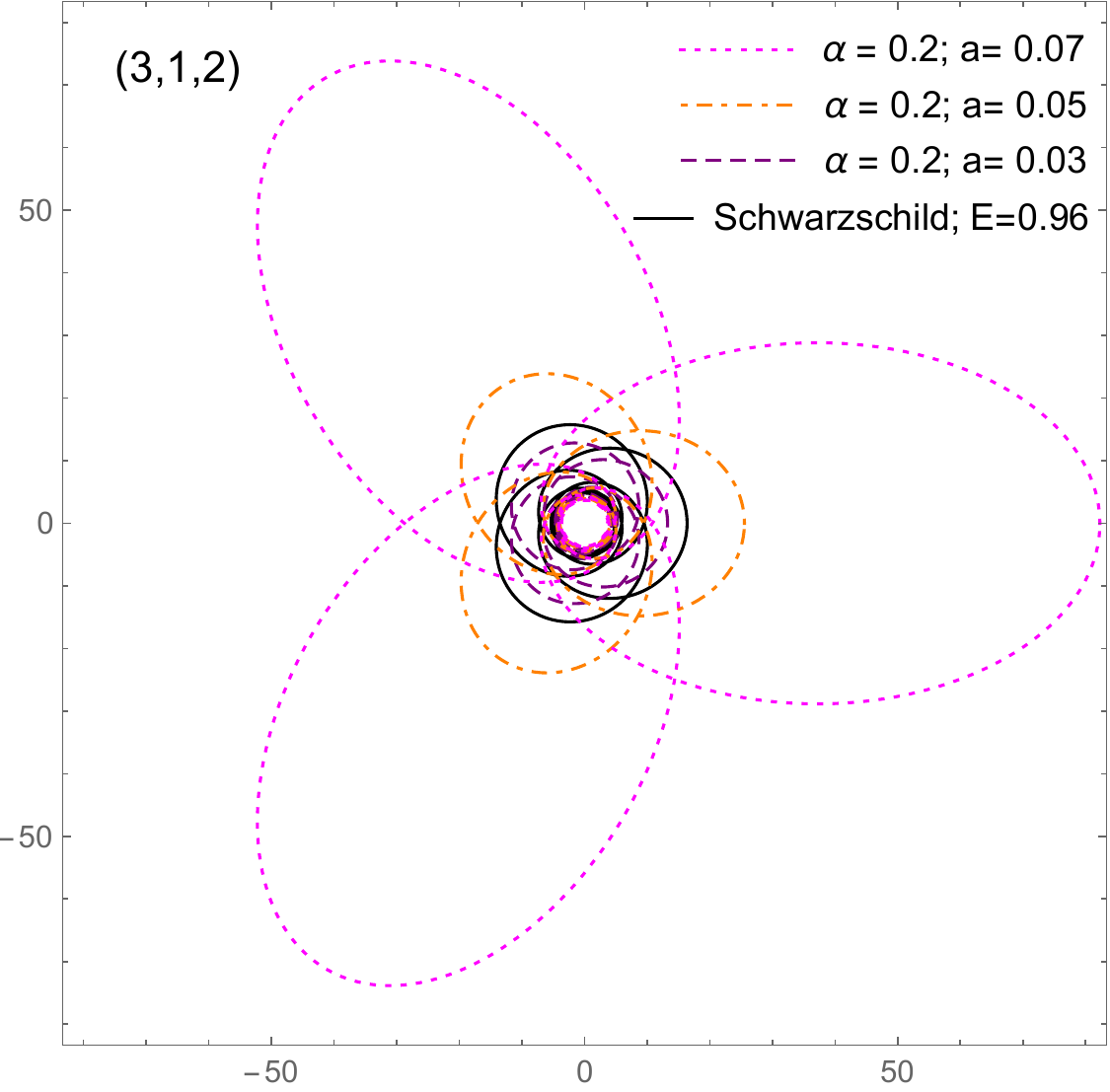}
		\end{subfigure}
		
		\vspace{0.5cm}

		\begin{subfigure}{0.3\textwidth}
			\centering
			\includegraphics[width=\linewidth, keepaspectratio]{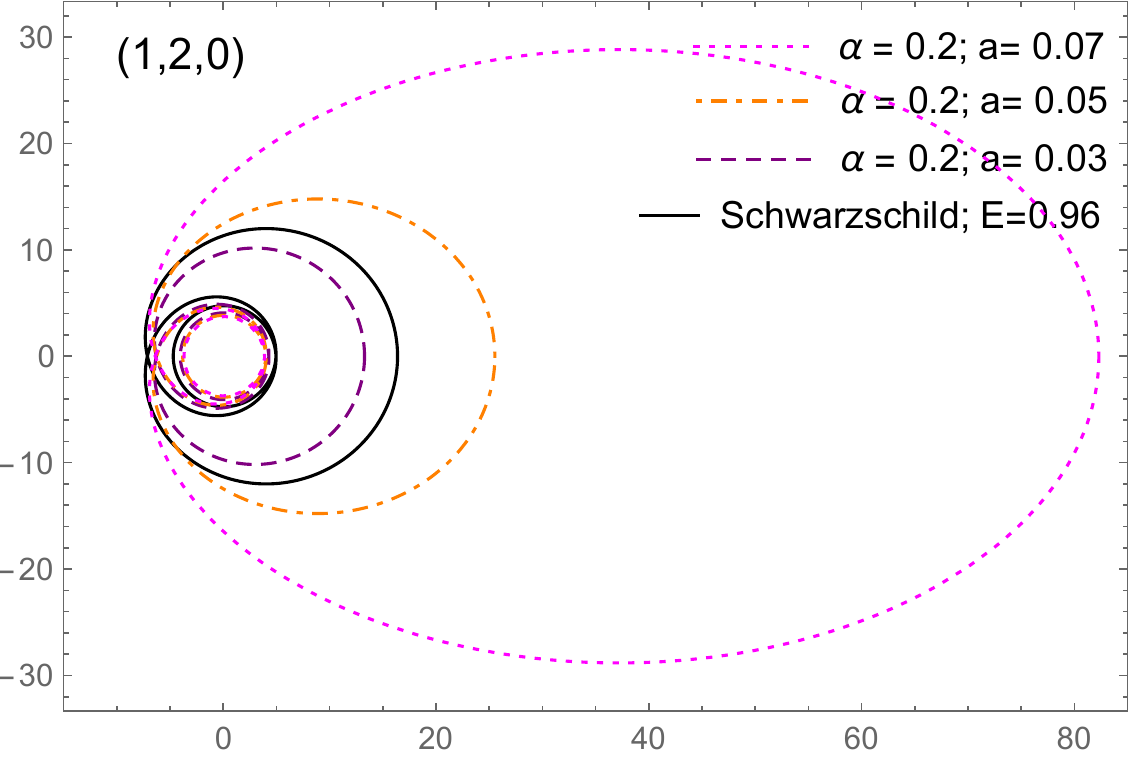}
		\end{subfigure}
		\hfill
		\begin{subfigure}{0.3\textwidth}
			\centering
			\includegraphics[width=\linewidth, keepaspectratio]{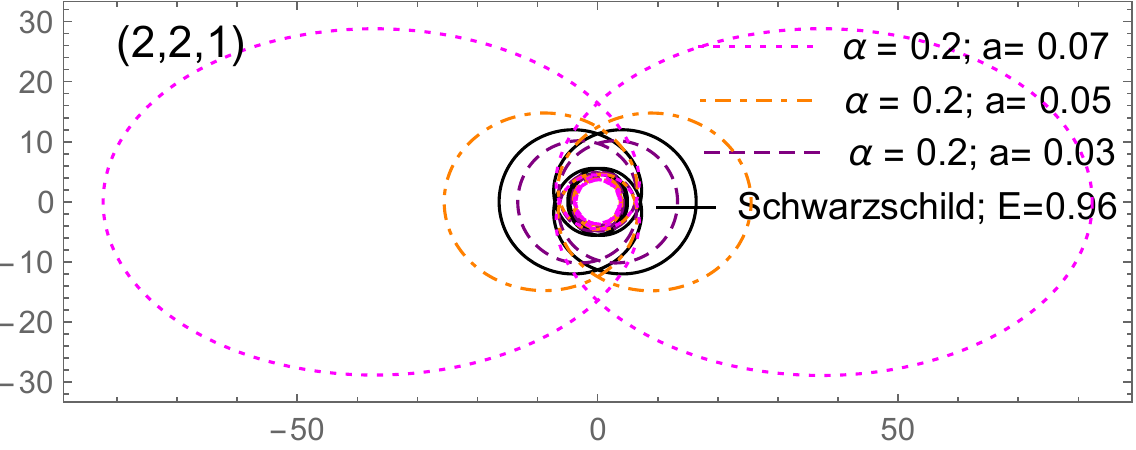}
		\end{subfigure}
		\hfill
		\begin{subfigure}{0.3\textwidth}
			\centering
			\includegraphics[width=\linewidth, keepaspectratio]{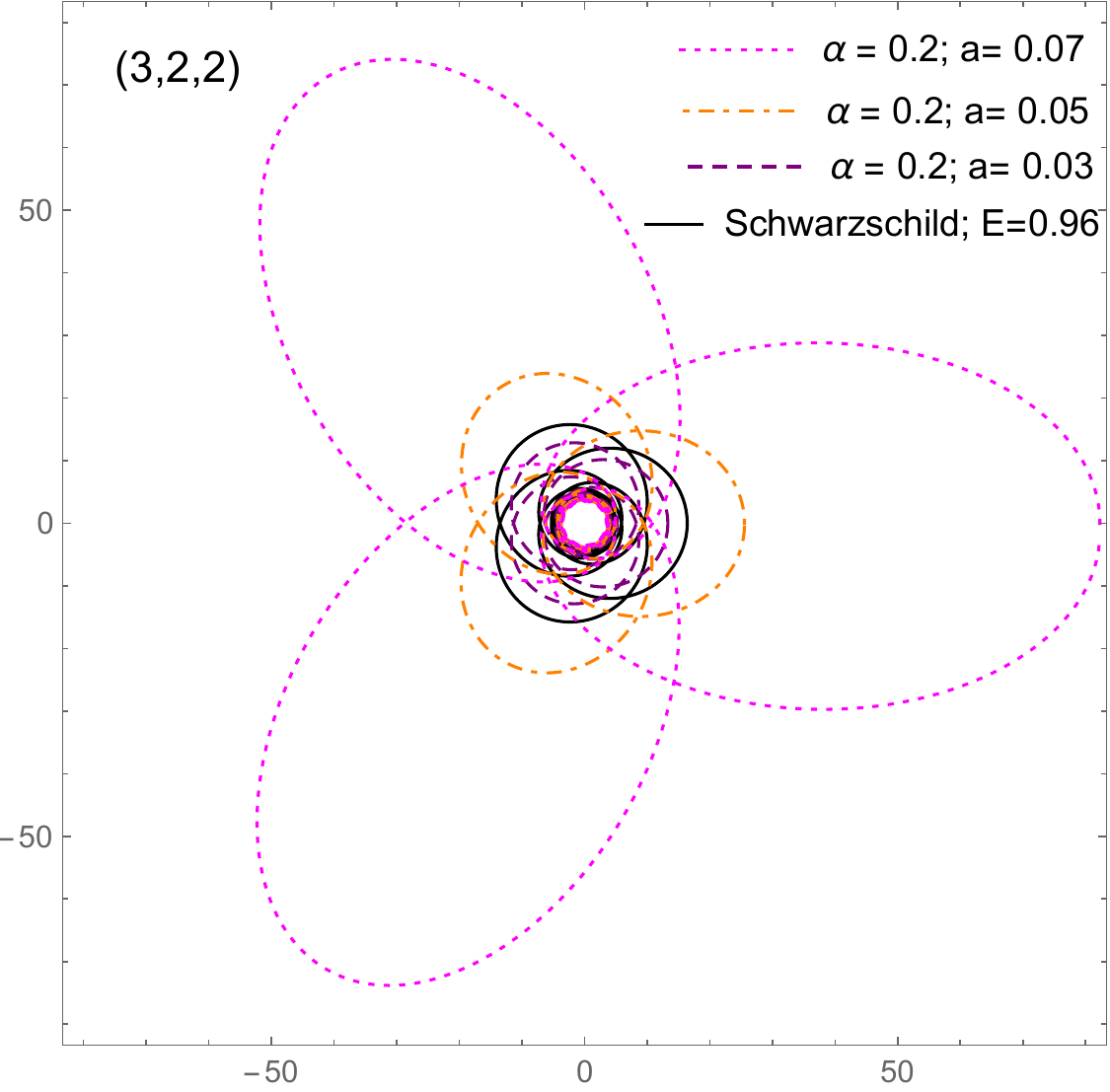}
		\end{subfigure}
		
		\caption{Periodic orbits labeled by $(z, w, v)$ around a Schwarzschild BH with a cloud of strings embedded in perfect fluid dark matter, shown for various values of the parameter $a$ with fixed energy $E = 0.96$ and $\alpha = 0.2$. The black curves correspond to the pure Schwarzschild case.}
		\label{orbitE}
	\end{figure*}

	We have calculated the properties of $(z, w, v)$-indexed periodic orbits in a Schwarzschild BH background modified by a string cloud and perfect fluid dark matter. Holding $\alpha = 0.2$ constant, we evaluated the energy $E$ (at fixed $L = (L_{\mathrm{MBO}} + L_{\mathrm{ISCO}})/2$) and the angular momentum $L$ (at fixed $E = 0.96$) for various values of the parameter $a$. As documented in Tables \ref{tabE} and~\ref{tabL}, our results demonstrate that an increase in the parameter $a$ leads to a decrease in the particle energy and an increase in the orbital angular momentum. Figures \ref{orbitL} and \ref{orbitE} display periodic orbits characterized by different indices $(z, w, v)$ in the spacetime of a Schwarzschild BH surrounded by a cloud of strings and perfect fluid dark matter. These orbits are computed for varying values of the parameter $a$ (with $\alpha$ fixed at $0.2$), under two conditions: fixed energy $E = 0.96$ or fixed angular momentum $L = (L_{\mathrm{MBO}} + L_{\mathrm{ISCO}})/2$. Several key features emerge from these plots: orbits with a larger zoom number $z$ exhibit richer structural complexity, while those with a larger whirl number $w$ undergo more revolutions around the central BH between successive apoapses.
	\begin{table*}[!htbp]  % 使用 table* 环境
		\setlength{\abovecaptionskip}{0.2cm}
		\setlength{\belowcaptionskip}{0.2cm}
		\centering
		\caption{The energy $E$ corresponding to periodic orbits characterized by $(z, w, v)$ for fixed angular momentum $L = \frac{1}{2}(L_{\mathrm{MBO}} + L_{\mathrm{ISCO}})$ and fixed dark matter parameter $\alpha = 0.2$.}
		\label{tabE}
		\resizebox{\textwidth}{!}{%  % 改为 columnwidth
			\begin{tabular}{lcccccccc}
				\hline 
				$\alpha$ & $a$ & $L$ & $E_{(1,1,0)}$ & $E_{(1,2,0)}$ & $E_{(2,1,1)}$ & $E_{(2,2,1)}$ & $E_{(3,1,2)}$ & $E_{(3,2,2)}$ \\
				\hline
				Schwarzschild &  & $ 3.7320508$ & $ 0.9654253$ & $ 0.9683828$ & $ 0.9680265$ & $ 0.9684343$ & $ 0.9682249$ & $  0.9684385$ \\
				\hline
				$0.2$ & $0.03$ & $ 2.8799500$ & $ 0.9678471$ & $ 0.9710367$ & $ 0.9706638$ & $ 0.9710885$ & $ 0.9708731$  & $   0.9710925$\\
				$0.2$ & $0.05$ & $ 2.9678154$ & $ 0.9621009$ & $ 0.9654625$ & $ 0.9650802$ & $ 0.9655132$ & $ 0.9652965$  & $   0.9655170$\\
				$0.2$ & $0.07$ & $ 3.0591250$ & $ 0.9561844$ & $ 0.9598148$ & $ 0.9594050$ & $ 0.9598674$ & $ 0.9596381$  & $   0.9598712$\\
				\hline 
			\end{tabular}%
		}
	\end{table*}
	
	\begin{table*}[!htbp]  % 使用 table* 环境
		\setlength{\abovecaptionskip}{0.2cm}
		\setlength{\belowcaptionskip}{0.2cm}
		\centering
		\caption{The orbital angular momenta $L$ corresponding to periodic orbits characterized by $(z, w, v)$, computed with the particle energy fixed at $E = 0.96$ and the dark matter parameter set to $\alpha = 0.2$.}
		\label{tabL}
		\resizebox{\textwidth}{!}{%  % 改为 columnwidth
			\begin{tabular}{lccccccc}
				\hline 
				$\alpha$ & $a$ & $L_{(1,1,0)}$ & $L_{(1,2,0)}$ & $L_{(2,1,1)}$ & $L_{(2,2,1)}$ & $L_{(3,1,2)}$ & $L_{(3,2,2)}$ \\
				\hline
				Schwarzschild &  & $  3.6835877$ & $  3.6534056$ & $  3.6575957$ & $  3.6527007$ & $  3.6553345$ & $   3.6526363$ \\
				\hline
				$0.2$ & $0.03$ & $  2.8201255$ & $  2.7915839$ & $  2.7956795$ & $  2.7908579$ & $  2.7934916$  & $    2.7907886$\\
				$0.2$ & $0.05$ & $  2.9519034$ & $  2.9253999$ & $  2.9286595$ & $  2.9249263$ & $  2.9268482$  & $    2.9248879$\\
				$0.2$ & $0.07$ & $  3.0888441$ & $  3.0605398$ & $  3.0636702$ & $  3.0601387$ & $  3.0618880$  & $    3.0601094$\\
				\hline 
			\end{tabular}%
		}
	\end{table*}
	
	\section{Gravitational waveforms from periodic orbits}
	\label{section4}
	EMRI systems serve as prime laboratories for probing strong-field gravity, as the emitted GWs carry distinct imprints of the background spacetime. Building upon the periodic orbital solutions obtained in the preceding section, we analyze the GW signals generated by a test particle moving in the vicinity of a Schwarzschild BH threaded by a cloud of strings and enveloped in dark matter. Within the framework of this EMRI model, the corresponding metric perturbation at quadratic order is given by the following expression \cite{Maselli:2021men,Liang:2022gdk}:
	\begin{equation}
		h_{ij}=\frac{4G_{\eta}M}{c^{4}D_{L}}\left(v_{i}v_{j}-\frac{Gm}{r}n_{i}n_{j}\right),\label{hij}
	\end{equation}
	In the above expression, $M$ denotes the mass of the supermassive BH, $m$ represents the mass of the orbiting particle, and $D_L$ corresponds to the luminosity distance of the EMRI system. Furthermore, $\eta = mM/(m + M)^2$ is defined as the symmetric mass ratio, $v$ denotes the relative velocity of the particle, and $n$ is the unit vector in the radial direction. To facilitate comparison with observational data, the gravitational wave solution derived in Eq. (\ref{hij}) is transformed into the detector's reference frame. This transformation yields the corresponding plus and cross polarization components, denoted as $h_{+}$ and $h_{\times}$, which are explicitly given by
	\begin{equation}
		h_{+}=-\frac{2\eta M}{c^{4}D_{L}}\frac{(GM)^{2}}{r}(1+\cos^{2}\iota)\cos(2\phi+2\omega),\label{hz}
	\end{equation}
	\begin{equation}
		h_{\times}=-\frac{4\eta M}{c^{4}D_{L}}\frac{(GM)^{2}}{r}\cos\iota\sin(2\phi+2\omega).\label{hc}
	\end{equation}
	Here, $\iota$ characterizes the orbital inclination, defined as the angle between the line of sight to the observer and the direction of the test particle's orbital angular momentum. Meanwhile, $\omega$ quantifies the longitude of pericenter, which specifies the orbital orientation of the closest approach.
	
\begin{figure*}[htbp]
		\centering
		\begin{subfigure}{\textwidth}
			\centering
			\includegraphics[width=0.6\linewidth]{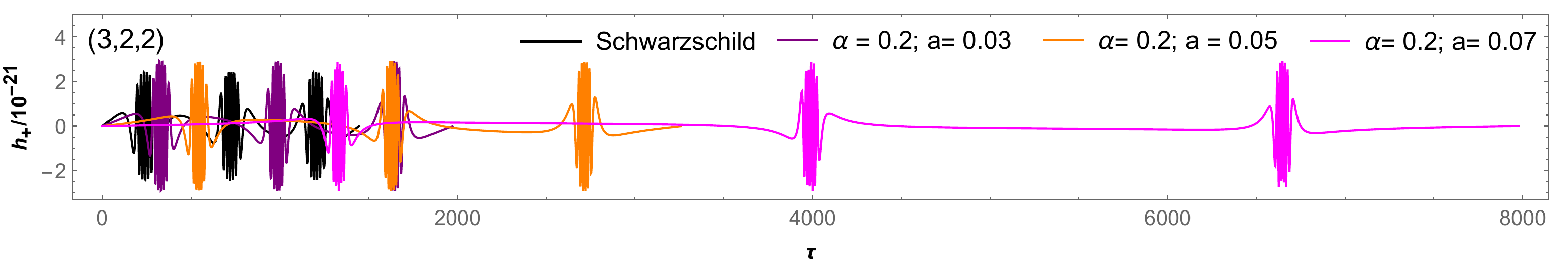}
		\end{subfigure}
		
		\vspace{0.5cm} % 调整上下间距
		
		\begin{subfigure}{\textwidth}
			\centering
			\includegraphics[width=0.6\linewidth]{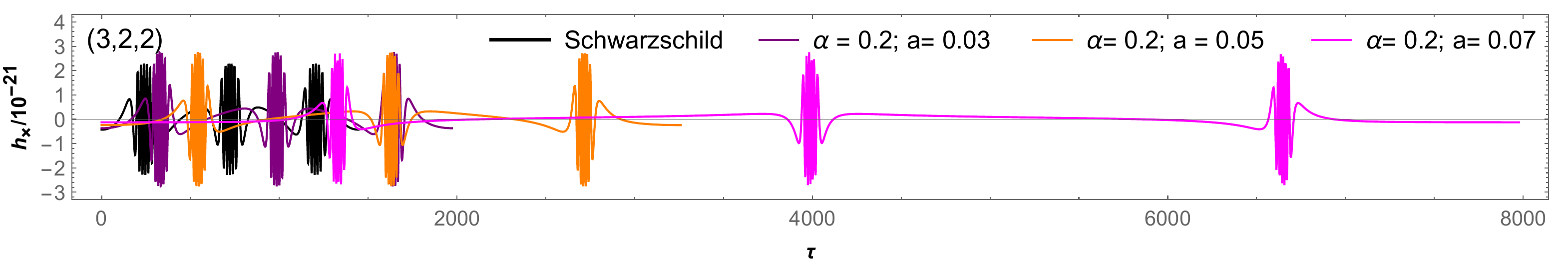}
		\end{subfigure}
		
		\caption{Gravitational waveforms emitted by a test particle of mass $m = 10\,M_\odot$ moving along $(3,2,2)$ periodic orbits around a Schwarzschild BH surrounded by a cloud of strings and embedded in perfect fluid dark matter. The black hole mass is $M = 10^{7}\,M_\odot$. The orbital angular momentum is fixed at $L = \tfrac{1}{2}(L_{\mathrm{MBO}} + L_{\mathrm{ISCO}})$, and the corresponding energy is determined by the orbit parameters.}
		\label{GW1}
	\end{figure*}

\begin{figure*}[htbp]
		\centering
		\begin{subfigure}{\textwidth}
			\centering
			\includegraphics[width=0.6\linewidth]{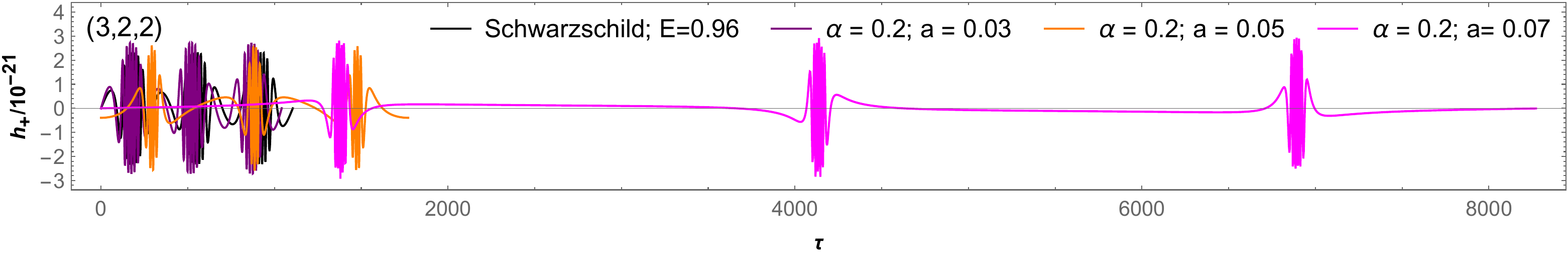}
		\end{subfigure}
		
		\vspace{0.5cm} % 调整上下间距
		
		\begin{subfigure}{\textwidth}
			\centering
			\includegraphics[width=0.6\linewidth]{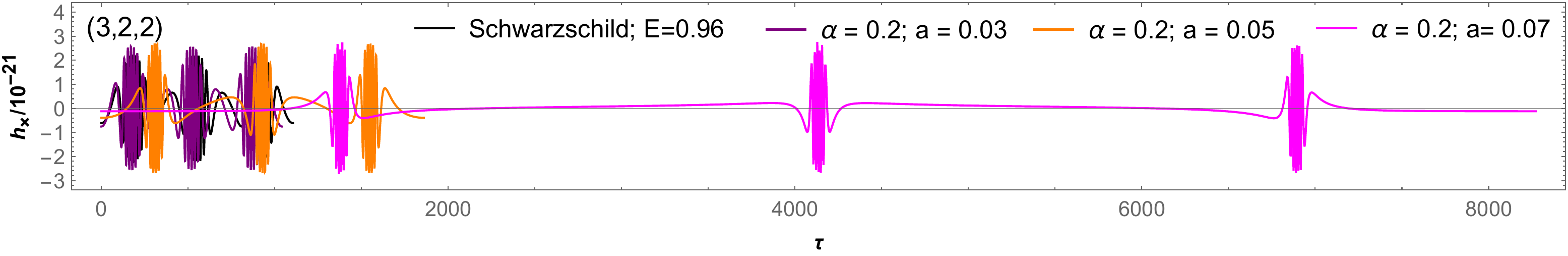}
		\end{subfigure}
		
		\caption{Gravitational waveforms emitted by a test particle of mass $m = 10\,M_\odot$ moving along $(3,2,2)$ periodic orbits around a Schwarzschild BH surrounded by a cloud of strings and embedded in perfect fluid dark matter. The black hole mass is $M = 10^{7}\,M_\odot$, and the particle energy is fixed at $E = 0.96$.}
		\label{GW2}
	\end{figure*}

	We investigate the influence of the parameters $a$ and $\alpha$ on gravitational waveforms by systematically varying them within $(3,2,2)$ periodic orbits, with the resultant waveforms shown in Figs. \ref{GW1} and \ref{GW2}. Our numerical simulations are based on the following fiducial EMRI system: $M = 10^7 M_\odot$, $m = 10 M_\odot$, $\iota = \pi/4$, $\omega = \pi/4$, and $D_L = 200\,\text{Mpc}$. As shown in the figures, increasing the string cloud parameter $a$ introduces a significant phase delay in the waveforms. Specifically, waveforms with lower $a$ values oscillate within a shorter time domain, while those with higher $a$ values extend to later time intervals. Notably, the black dashed curve serves as a benchmark, representing the pure Schwarzschild BH case.
	
	\section{Conclusion} 
	\label{section5}
	In this paper, we have investigated the geodesic structure and gravitational wave phenomenology in the spacetime of a Schwarzschild BH surrounded by a cloud of strings and embedded in perfect fluid dark matter. By introducing the parameters $a$ and $\alpha$ to characterize the string cloud and dark matter medium, respectively, we have systematically analyzed the behavior of test particle orbits, focusing on key relativistic features such as the ISCO and MBO.
	
	Our analysis reveals that the presence of the string cloud and dark matter significantly modifies the orbital dynamics. Specifically, we find that the orbital radius and angular momentum of circular orbits increase with $a$ but decrease with $\alpha$, whereas the energy shows an inverse dependence. These shifts directly impact the location of the ISCO and MBO, altering the efficiency of energy release and the inner edge of accretion disks. Moreover, the study of periodic orbits—classified by rational numbers $q$—demonstrates rich trajectory structures whose properties are sensitive to environmental parameters.
	
	Crucially, the resulting gravitational waveforms exhibit distinctive signatures: an increase in $a$ leads to a notable phase delay and a time-domain shift of the waveform envelope, with high-parameter waveforms appearing at later times and displaying altered amplitudes. These deviations from the standard Schwarzschild case are substantial enough to allow clear discrimination between BHs in such exotic environments and vacuum ones.
	
	Therefore, our results suggest that future high-precision gravitational wave observations can serve as a powerful tool for probing not only the strong-field regime of gravity but also the nature of dark matter and topological defects like cosmic strings in the vicinity of BHs. The distinct waveform morphologies identified here provide a viable pathway for constraining the parameters $a$ and $\alpha$ through observational data, opening a new window into the astrophysics of BH environments.
	
	\begin{acknowledgments}
		This study is supported in part by National Natural Science Foundation of China (Grant
		No. 12333008) and Guizhou Provincial Major Scientific and Technological Program XKBF (2025)010.
	\end{acknowledgments}

\end{document}